\documentclass[12pt,a4paper]{article}
\usepackage{type1cm,amsmath,graphicx,indentfirst}
\usepackage[psamsfonts]{amssymb}
\numberwithin{equation}{section}

\def\slash#1{\not\!#1}
\def\slashb#1{\not\!\!#1}

\begin{document}
\begin{titlepage}

 \renewcommand{\thefootnote}{\fnsymbol{footnote}}
\begin{flushright}
 \begin{tabular}{l}
 \end{tabular}
\end{flushright}

 \vfill
 \begin{center}


\noindent{\large \textbf{Higher spin AdS$_3$ supergravity and its dual CFT
}}\\
\vspace{1.5cm}

\noindent{ Thomas Creutzig,$^{a}$\footnote{E-mail: tcreutzig@mathematik.tu-darmstadt.de} Yasuaki Hikida$^b$\footnote{E-mail:
hikida@phys-h.keio.ac.jp} and Peter B. R\o nne$^{c}$\footnote{E-mail: peter.roenne@uni-koeln.de}}
\bigskip

 \vskip .6 truecm
\centerline{\it $^a$Fachbereich Mathematik,
Technische Universit\"{a}t Darmstadt,}
\centerline{\it Schlo\ss gartenstr. 7
64289 Darmstadt, Germany}
\medskip
\centerline{\it $^b$Department of Physics, and Research and Education
Center for Natural Sciences,}
\centerline{\it  Keio University, Hiyoshi, Yokohama 223-8521, Japan}
\medskip
\centerline{\it $^c$Institut f\"{u}r Theoretische Physik, Universit\"{a}t zu K\"{o}ln,
} \centerline{\it
Z\"{u}lpicher Stra{\ss}e 77, 50937 Cologne, Germany}
 \vskip .4 truecm

 \end{center}

 \vfill
\vskip 0.5 truecm

\begin{abstract}

Vasiliev's higher spin supergravity theory on three dimensional anti-de Sitter
space is studied and, in particular,
the partition function is computed at one loop level.
The dual conformal field theory is proposed to be the ${\cal N}=(2,2)$ $\mathbb C$P$^N$
Kazama-Suzuki model in two dimensions.
The proposal is based on symmetry considerations and comparison of the bulk partition function with the conformal field theory.
Our findings suggest that the theory is strong-weak self-dual.

\end{abstract}
\vfill
\vskip 0.5 truecm

\setcounter{footnote}{0}
\renewcommand{\thefootnote}{\arabic{footnote}}
\end{titlepage}

\newpage

\tableofcontents

\section{Introduction}

In \cite{GG} Gaberdiel and Gopakumar suggested a holographic duality relating a bosonic higher spin theory on three dimensional anti-de Sitter space
(AdS$_3)$ \cite{PV1,PV2} to a large $N$ limit of a certain two dimensional
conformal field theory (CFT$_2$).
This is a three dimensional version of the duality proposed by Klebanov and
Polyakov \cite{KP} which states that a Vasiliev's higher spin
theory on AdS$_4$ \cite{Vasiliev:2003ev} is dual to the three dimensional
O$(N)$ vector model. These can be interpreted as simplified versions of
AdS/CFT correspondence \cite{Maldacena:1997re}, and they provide a good terrain to test and understand holography deeper.
In particular, holographic dualities between higher spin gravity on AdS$_3$ and
a large $N$ limit of a two-dimensional conformal field theory have the advantage
that both theories can be fairly well studied.
In this paper, we would like to propose a supersymmetric version of the
duality in \cite{GG}.
In general, it is a natural and good idea to add supersymmetry as
supersymmetric theories are often better behaved than their bosonic counterparts.

The higher spin gravity used in \cite{GG} is a truncated version of Prokushkin and Vasiliev's 3d supersymmetric theory \cite{PV1,PV2} with infinite towers of higher spin fields coupled to massive matter.
The truncated version contains one infinite tower of higher spin fields with symmetry algebra hs[$\lambda$], $0\leq \lambda\leq 1$, and two complex scalar particles having the same mass which is related to the parameter $\lambda$ by
\begin{align}\label{eq:massminus}
 ( M^B_-)^2 = - 1 + \lambda  ^2 \ .
\end{align}
Without coupling to matter, the higher spin algebra can be cut off at some finite spin $N$.
The theory is described by the 3d Chern-Simons theory based on two copies of sl$(N)$  \cite{Blencowe},
and the asymptotic symmetry near the boundary is found to be the
${\cal W}_N$ algebra \cite{HR,CFPT,GH,CFP}.
This is in turn the symmetry of the ${\cal W}_N$ minimal coset model
\begin{align}\label{eq:GGcoset}
 \frac{\text{SU}(N)_k \times \text{SU}(N)_1}{\text{SU}(N)_{k+1}} ~.
\end{align}
With the help of this fact, it was proposed in \cite{GG}
that the dual CFT is given by taking ``the 't Hooft limit'' of the ${\cal W}_N$ minimal coset
where we are send   $N,k \to \infty$, but keep the 't Hooft coupling
\begin{align}\label{eq:coupling}
 \lambda = \frac{N}{k+N}
\end{align}
fixed.

Besides this check via asymptotic symmetry, the conjectured duality was also checked by seeing that the RG flow of the coset theory could be reproduced by the bulk theory \cite{GG}. And even more impressing, it is possible to check that the partition function of the coset CFT matches the one loop determinant of the bulk theory. The one loop determinant was calculated in \cite{GGS,GG} and contains the vacuum character of the ${\cal W}$ algebra as a factor. The remaining part corresponds to the massive matter and is captured on the CFT side as (multi-particle) finite fusion products of the four representations  dual to the massive matter \cite{GGHR}. The conjecture is that those states which are created via fusion, but do not compare to the bulk spectrum, decouple in the 't Hooft limit. There has been further checks considering three- and four-point functions \cite{Chang:2011mz,Papadodimas:2011pf,Ahn:2011by}, but they have revealed some unsolved puzzles relating to the decoupling of the remaining states.

The large $N$ limits of other cosets have been considered in \cite{Kiritsis:2010xc}, and the case where one considers ${\cal W} B_N$ and ${\cal W} D_N$ algebras instead of the above used ${\cal W} A_N$ algebras are investigated in \cite{Ahn, Gaberdiel:2011nt}.
Recently, an attempt to construct dS$_3/$CFT$_2$ is made in \cite{Ouyang:2011fs}.
Moreover, the gravity duals of minimal models without the large $N$ limit are proposed in \cite{Ising}.
In this paper we would like to go in another direction and construct an $\mathcal N=2$ supersymmetric version of
the duality proposed in \cite{GG}.

Let us first consider the gravity theory without matter and with the spin
truncated $s \leq N+1$.
We do this by considering
the Chern-Simons theory based on the copies of the supergroup sl$(N+1|N)$ instead of sl$(N)$, as was also suggested by Ahn \cite{Ahn}.
The supergravity algebra is found for $N=1$ where $\text{sl}(2|1)\simeq \text{osp}(2|2)$ and this theory was first studied in \cite{AT}, see also \cite{HMS} in our context.
We now add massive matter and will have to consider an infinite tower of higher spins.
The bulk theory should be the untruncated theory given in \cite{PV1,PV2}. The higher spins generate a symmetry algebra which we denote
by shs[$\lambda$].%
\footnote{This superalgebra is the ${\cal N}=2$ supersymmetric version of hs$[\lambda]$. See appendix \ref{hsa} for details.}
The matter forms a 3d $\mathcal N=2$ massive hypermultiplet  which contains two extra scalars with mass
\begin{align}\label{}
 ( M^B_+ )^2 = - 1 + (\lambda - 1 )^2 ~,
\end{align}
in addition to the two scalars with mass given in \eqref{eq:massminus}.
Further we have four Dirac fermions with mass
\begin{align}\label{}
 (M^F_\pm)^2 = (\lambda - \tfrac{1}{2})^2 ~.
\end{align}

For the dual 2d CFT we consider the asymptotic symmetry of the finite spin bulk theories. For Chern-Simons theory with sl$(N+1|N)$ symmetry,
the asymptotic symmetry is the ${\cal N}=2$ super ${\cal W}_{N+1}$ algebra.
A minimal model with symmetry given by the ${\cal N}=2$ super ${\cal W}_{N+1}$ algebra
is  the $\mathbb C$P$^N$ Kazama-Suzuki coset
\cite{KS}
\begin{align}\label{scoset}
 \frac{\text{SU}(N+1)_k \times \text{SO}(2 N)_1 }{\text{SU}(N)_{k+1} \times \text{U}(1)_{N(N+1)(k+N+1)} } ~.
\end{align}
From this consideration,
we propose that the dual CFT of the higher spin supergravity theory is given by
the 't Hooft limit \eqref{eq:coupling} of the super coset \eqref{scoset}.
The limit must take this form since the case in \cite{GG} is a truncation of this theory.
Interestingly, the coset possesses a level-rank duality \cite{Naculich:1997ic} which
in the 't Hooft limit looks like a strong-weak duality exchanging the couplings
\begin{equation}
\lambda \ \longleftrightarrow \ 1-\lambda\, .
\end{equation}
In the higher spin bulk theory we see similar features.

The organization of this paper is as follows.
We start by introducing higher spin supergravity in section 2.
The higher spin theory is a large $N$ limit of SL$(N+1|N)$ $\otimes$ SL$(N+1|N)$ Chern-Simons theory.
We then discuss the asymptotic symmetry of the theory.
In section 3 and 4 we compute the gravity partition function. Here section 4 is the detailed computation of the one loop determinant for higher spin fermionic
particles. The result is summarized in \eqref{hsf} and actually
this is one of the main findings of this paper.
In section 5 we compare the bulk partition function with the dual CFT. We find that the vacuum character of the $\mathcal N=(2,2)$ ${\cal W}$ algebra
agrees with the massless part of the bulk partition function.
We then use supersymmetry considerations as well as level-rank duality to observe that the bulk partition function is indeed
a very reasonable candidate for the CFT partition function. Finally, we suggest four states corresponding to the massive bulk matter. The characters of these states are related to the known characters of the coset considered by Gaberdiel and Gopakumar \eqref{eq:GGcoset} and we calculate them to low order and make a comparison with the bulk partition function. 
We then conclude with future directions.
In appendix \ref{hsa} the infinite dimensional algebras
hs$[\lambda]$ and shs$[\lambda]$ are reviewed, and in appendix \ref{heat}
the heat kernel method of \cite{GMY,DGG} is
explained with two important examples.

\section{Higher spin AdS$_3$ supergravity}

Higher spin AdS$_3$ supergravities can be defined as supergroup Chern-Simons theories \cite{AT}. We are going to study
$\text{SL}(N+1|N) \otimes \text{SL}(N+1|N) $ Chern-Simons theory,
and its asymptotic symmetry around the AdS boundary.
The higher spin sector of the Vasiliev's theory \cite{PV1,PV2}
can be described by the $\text{shs}[\lambda] \otimes \text{shs}[\lambda] $
Chern-Simons theory \cite{Blencowe} where $\text{shs}[\lambda]$ can be
thought of as a large $N$ limit, or analytical continuation, of $\text{sl}(N+1|N)$, see appendix \ref{hsa}.
In the next subsection, we introduce the Chern-Simons theory with
$\text{sl}(N+1|N)$ symmetry and study its spectrum.
In subsection \ref{asymsym} we find that the  asymptotic symmetry
near the boundary is ${\cal N}=(2,2)$ super ${\cal W}_{N+1}$ algebra.

\subsection{Chern-Simons formulation}

In 2+1 dimensions, Einstein gravity with negative cosmological constant is known to be equivalent to
Chern-Simons theory with action \cite{AT}
\begin{align}
 S = S_\text{CS} [A] - S_\text{CS} [\tilde A] ~,
 \label{CSaction}
\end{align}
where
\begin{align}
 S_\text{CS} [ A ] = \frac{\hat k}{4 \pi} \int \text{tr}
\left( A \wedge dA + \frac{2}{3} A \wedge A \wedge A \right) ~.
\end{align}
The gauge fields $A , \tilde A$ take values in
$\text{sl}(2) \oplus \text{sl}(2)$.

For the moment, we focus on the part corresponding to $A$.
Denoting $A = A_\mu ^a J_a d x^\mu$, where $J_a$ $(a =1,2,3)$ generates sl$(2)$
algebra, the parameter $\hat k$ is
related to the Newton constant as
\begin{align}
 \hat k = \frac{\ell}{4 G}  \qquad \textrm{with}\qquad \text{tr} ( J_a J_b ) = \frac{1}{2} \delta_{a,b} ~.
\end{align}
Here $\ell$ is the AdS radius and it will be set to one, $\ell=1$, in the following.
The theory is invariant under the gauge transformations
\begin{align}
 \delta A = d \lambda + [A , \lambda] ~,
 \qquad \delta \tilde A = d \tilde \lambda + [\tilde A , \tilde \lambda] ~.
\label{gaugetrans}
\end{align}
In order to relate to the pure gravity theory in the first order formulation,
we may combine gauge fields as
\begin{align}
 e = \frac{1}{2} (A - \tilde A) ~, \qquad
\omega =   \frac{1}{2} (A + \tilde A) ~,
\end{align}
where $e_\mu^a$ is the dreibein and
$\omega_{\mu,a,b} = \frac12 \epsilon_{abc} \omega_{\mu}^c$ is the
spin connection.

In order to supersymmetrize the gravity theory, we have to replace
the Lie algebra by the Lie superalgebra along with the supertrace instead of the ordinary trace.
A supergravity theory may be defined by
$\text{OSP}(p|2) \otimes \text{OSP}(q|2)$ Chern-Simons theory \cite{AT}.
The theory has ${\cal N}= p + q$ supersymmetry, and
$p$ and $q$ gravitini have the opposite signature of "mass"
induced by the curvature of AdS space.
These fields also couple with
$\text{O}(q) \otimes \text{O}(p)$ Chern-Simons theory.
More generic cases are considered in \cite{HMS}, where they use
a type of supergroup with bosonic subgroup of the form
$\text{SL(2)} \otimes G$.

In the bosonic case,
the generalization to a higher spin gravity theory can be obtained by
replacing the algebras with $\text{sl}(N)$ or
$\text{hs}[\lambda]$ \cite{Blencowe}.
The $\text{SL}(N) \otimes \text{SL}(N)$ Chern-Simons theory includes
higher spin gauge fields up to spin $s \leq N$.
For recent developments, see \cite{HR,CFPT,CFP}.
In order to construct a higher spin AdS$_3$ supergravity theory,
we use the Lie superalgebra
$\text{sl}(N+1|N)$ or the infinite dimensional superalgebra
$\text{shs}[\lambda]$.
The gravity theory we would like to consider is the Vasiliev's theory
\cite{PV1,PV2}, and its massless part may be described by
Chern-Simons theory with $\text{shs}[\lambda]$ symmetry. The theory also includes
massive scalars and spin $1/2$ spinors coupled with the higher spin fields.
In this section we only consider the massless part.

As in \cite{CFP}
we decompose the $\text{sl}(N+1|N)$ element in terms of
a $\text{sl}(2)$ subalgebra as follows (see (29) of \cite{FL} for instance)
\begin{align}
 \text{sl}(N+1|N) = \text{sl}(2)   \oplus
 \left( \bigoplus _{s=3}^{N+1} \text{g}^{(s)} \right) \oplus
 \left( \bigoplus _{s=1}^N \text{g}^{(s)} \right) \oplus
 2  \times \left( \bigoplus _{s=1}^{N+1}  \text{g}^{(s+\frac12)} \right) ~.
 \label{slexp}
\end{align}
For $\text{shs}[\lambda]$ we can naively take the $N \to \infty$ for the
upper bound of the product.
This is the principal sl(2) embedding used for the Hamiltonian reduction of sl$(N+1|N)$
WZNW model to obtain ${\cal N}=2$ super Toda theory
(see, for instance, \cite{W-alg}). The first sl(2) may be referred to as
the "gravitational" one since it is the usual spin 2 gravity sector
introduced in the beginning of this section.
The other part $\text{g}^{(s)}$ is in the $2s-1$ dimensional representation of
sl(2). One merit of this decomposition is that the elements of integer $s$ part
are commuting variables and the elements of half-integer $s$ part
are anti-commuting variables. In other words, the spin-statistic relation holds in this case since $s$ is the spin for the Lorentz transformation.

We denote the generators of the Lie superalgebra as (see appendix \ref{hsa})
\begin{align}
V_n^{(s)+} ~ (s=2,3,\cdots) ~, \qquad V_n^{(s)-} ~ ( s=1,2,\cdots) ~, \qquad
F^{(s)\pm}_r ~ (s = 1, 2,\cdots)
\label{sslgen}
\end{align}
with $|n| \leq s-1, |r| \leq s - 1/2$.
The first two sets of generators are Grassmann even and the last one is Grassmann odd.
Here $V^{(2)+}_n$ $(n=0,\pm 1)$ are the generators of gravitational sl(2).
The gauge fields $A, \tilde A$ are expanded as
\begin{align}
\label{Adec}
 A &= \sum_{n,s, a = \pm}  \phi_{\mu, a}^{(s) n} V_n^{(s)a } d x^\mu
+ \sum_{n,r, a = \pm} \psi_{\mu, a}^{(s)r} F^{(s)a}_r  d x^\mu~, \\
 \tilde A &= \sum_{n,s, a = \pm}  \tilde \phi_{\mu, a}^{(s) n} V_n^{(s)a } d x^\mu
+ \sum_{n,r, a = \pm} \tilde \psi_{\mu, a}^{(s)r} F^{(s)a}_r  d x^\mu~.
\nonumber
\end{align}
Namely, there are higher spin gauge fields
\begin{align}
\label{bosondeg}
 e_{\mu, a}^{(s) n} = \frac{1}{2} (
 \phi_{\mu, a}^{(s) n} - \tilde  \phi_{\mu, a}^{(s) n} ) ~, \qquad
  \omega_{\mu, a}^{(s) n} = \frac{1}{2} (
 \phi_{\mu, a}^{(s) n} + \tilde  \phi_{\mu, a}^{(s) n} )   ~,
\end{align}
with $|n| \leq s-1$ for the bosonic part.
Here $s = 2,3, \cdots$ for $a=+$, and the $s=2$ part is the
gravitational sl(2) sub-sector. Moreover, $s = 1,2, \cdots$ for $a=-$.
Further, there are fermionic higher spin gauge fields
\begin{align}
\label{fermiondeg}
\psi^{(s)r}_{\mu. \pm} ~ , \qquad \tilde \psi^{(s)r}_{\mu. \pm} ~,
\end{align}
with $s = 1,2, \cdots$ and $|r| \leq s - 1/2$.
Following \cite{AT} the fields $\psi^{(s)r}_{\mu. \pm}$ and
$\tilde \psi^{(s)r}_{\mu. \pm}$ should have the opposite signature of "mass"
from AdS curvature.
The gauge symmetry is generated by the transformation \eqref{gaugetrans}
but now the fields $A , \tilde A$ takes values in the Lie superalgebra
$\text{sl}(N+1|N)$ or $\text{shs}[\lambda]$.

\subsection{Asymptotic symmetry}
\label{asymsym}

We would like to study the asymptotic symmetry of the Chern-Simons theory
near the boundary of AdS space.
We consider a space which is a product of a disk and the time direction, and their coordinates are $(\rho, \theta)$
and $t$. Here the radial coordinate is $\rho$ and the boundary is
at $\rho \to \infty$. The other coordinates $(t,\theta)$ are the boundary coordinates.
Gauge transformations allow us to set
\begin{align}
 A_+ = e^{- \rho V_0^{(2)+}} a (t + \theta) e^{ \rho V_0^{(2)+}} ~,
 \qquad A_- = 0 ~, \qquad
 A_\rho = e^{- \rho V_0^{(2)+}} \partial_\rho e^{ \rho V_0^{(2)+}} ~,
\end{align}
where $A_\pm = A_\theta \pm A_t$.
Here $a(t + \theta)$ is arbitrary and it can be expanded in the basis
\eqref{sslgen}.
For the application to the AdS/CFT correspondence, we
should assign a boundary condition corresponding to the asymptotical
AdS space. After assigning the condition, we can set \cite{HR,CFPT,CFP}
\begin{align}
\label{AdSgauge}
 a (t + \theta) = V^{(2)+}_1 &+ \sum_{s \geq 2} L^+_s (t + \theta )
V^{(s)+}_{-s+1} + \sum_{s \geq 1} L^-_s (t + \theta )
V^{(s)-}_{-s+1} \\ &+ \sum_{s \geq 1} G^+_s (t + \theta )
F^{(s)+}_{-s+\frac{1}{2}} + \sum_{s \geq 1} G^-_s (t + \theta )
F^{(s)-}_{-s+\frac{1}{2}} ~. \nonumber
\end{align}

The residual gauge transformation preserving the gauge fixing condition is
given by
\begin{align}
 \Lambda (t+\theta )
 =  e^{- \rho V_0^{(2)+}} \lambda (t + \theta) e^{ \rho V_0^{(2)+}} ~,
\end{align}
which leads to
\begin{align}
 \delta_\lambda a (\theta ) = \partial_\theta \lambda (\theta )
+ [a (\theta) , \lambda (\theta )]
\label{deltalambda}
\end{align}
at fixed time $t$. Here $\lambda (\theta)$ can be expanded in terms of
\eqref{sslgen}. In the presence of boundary, the gauge transformation
does not always generate a physically equivalent state.
In fact, the gauge transformation with $\lambda (\theta)$ not
vanishing at the boundary generates physical symmetries.
These symmetries are generated by the boundary charges
\begin{align}
 Q (\lambda ) = - \frac{k}{2 \pi} \int d \theta \, \text{str} \, (\lambda (\theta ) a (\theta ))
 ~.
\end{align}
{}From the original Chern-Simons action, we can obtain the
classical Poisson brackets.
Now that we have assigned a gauge fixing and AdS boundary condition, the
phase space is reduced from the original one.
Since the classical Poisson brackets for the reduced phase space are given by
\begin{align}
 \delta_\lambda a (\theta ) =
 \{ Q (\lambda ) , a (\theta) \} ~,
\end{align}
we have
\begin{align}
 \{ Q (\lambda ) , Q(\eta) \} =
  - \frac{k}{2 \pi} \int d \theta \, \text{str} \, (\lambda (\theta ) \delta_\lambda a (\theta )) ~.
\end{align}
Computing the Poisson brackets explicitly, we can obtain the classical Poisson structure of $L^\pm_s (\theta)$ and $G^\pm_r(\theta)$.
In practice, we compute \eqref{deltalambda} with arbitrary $\lambda (\theta)$.
In general $a(\theta) + \delta a (\theta)$ is not of the form
\eqref{AdSgauge}, and the restriction to this form gives constraints
on  $\lambda (\theta)$. After solving the conditions, we can in principle obtain
the classical Poisson structure and from it we can read off the asymptotic
symmetry of the gravity theory.%
\footnote{Since we are dealing with normalizable modes for the massive scalars and
spin $1/2$ spinors, these fields do not contribute to the asymptotic symmetry.
Therefore, we just need to focus on the massless part.}

It is actually pointed out in \cite{CFPT,CFP} that the condition coming from imposing the
AdS boundary condition \eqref{AdSgauge}
is equivalent to the Drinfeld-Sokolov reduction
of the corresponding current algebra. Therefore,
we expect that for the case with $\text{sl}(N+1|N)$
the classical Poisson structure is the one for
${\cal N}=2$ ${\cal W}_{N+1}$ algebra since we use the $\text{sl}(2)$
principal embedding used for the Hamiltonian reduction \cite{W-alg}.
As pointed out in \cite{CFPT}, the higher spin part does not modify the
sl(2) subsector, therefore the asymptotic symmetry algebra includes the
Virasoro algebra with the central charge
\begin{align}
 c = 12 \hat k \, \text{str}  (V_0^{(2)+} V_0^{(2)+}) = \frac{3l}{2 G} ~.
\end{align}
For shs[$\lambda$], there might be subtleties since it is infinite
dimensional. The detailed investigation is left for future work.

\section{Supergravity partition function}

In this subsection
we compute the partition function of Vasiliev's theory with
symmetry $\text{shs}[\lambda] \otimes \text{shs}[\lambda]$.
The partition function is important as it allows us to read off the spectrum of the theory.
The partition function of the bosonic sub-sector was computed in
\cite{GG} (see also \cite{GMY,GGS}), and the computation for the
fermionic sub-sector is new.

\subsection{Higher spin gauge fields}

We start from the massless higher spin gauge fields.
In the Chern-Simons formulation, the gauge field $A$ can
be expanded as \eqref{Adec}, and the bosonic fields
and fermionic fields are as in \eqref{bosondeg} and
\eqref{fermiondeg}.
We want to compute the one loop contribution of these fields to the
thermal partition function.
Thus we consider the thermal AdS space, where the boundary is a torus with
modular parameter $q = \exp (2\pi i \tau)$.

The one loop determinants of spin $s \geq 2$ gauge fields were
computed in \cite{GGS} utilizing the heat kernel method
\cite{GMY,DGG}. See also appendix \ref{heat}.
Our theory is in the first order formulation, and by integrating
over the connection-like fields $\omega^{(s)n}_{\mu,\pm}$ in
\eqref{bosondeg} we obtain the action for $e^{(s)n}_{\mu,\pm}$
in \eqref{bosondeg} with second order derivatives.
In order to map from the frame-like formulation to the metric-like
formulation, we have to change basis.
Notice that
$e^{(s)n}_{\mu,\pm}$ with $|n| \leq s-1$ can be described by
symmetric traceless expressions as
$e_{\mu a_1 \cdots  a_{s-1}}^{\pm }$ with $a_i = 1,2,3$.
Then we define gauge fields with higher spin $s$ as
\begin{align}
 \varphi^\pm_{\mu_1  \cdots  \mu_{s}}
  = \frac{1}{s} \bar e_{(\mu_1}^{~~~a_1} \cdots \bar e_{\mu_{s-1}}^{~~~a_{s-1}}
  e_{\mu_s )  a_1 \cdots a_{s-1}}^\pm ~.
  \label{bosonbasis}
\end{align}
Here $\bar e_\mu^{~a}$ is the background dreibein on the AdS space, and
the parenthesis denotes the complete symmetrization of the indices enclosed.
The free action for these fields on the AdS space was obtained in \cite{FronsdaldS}
(see also, e.g., \cite{Campoleoni}).
Using the action, the partition function at one loop level was computed
in \cite{GGS}. For spin $s \geq 2$ gauge fields, it is obtained as
\begin{align}
Z^{(s)}_B = \frac{\text{det}^{\frac12} \left( - \Delta + s (s-1) \right)^\text{TT}_{(s-1)}}{\text{det}^{\frac12} \left( - \Delta + s (s-3) \right)^\text{TT}_{(s)}} ~.
\end{align}
As the subscripts indicate, the Laplacian $\Delta$ on the AdS space acts only
for the transverse traceless components of spin $s$ and $s-1$ gauge fields.
Applying the formula \eqref{hkdet}, the one loop determinants can be computed.
The results are
\begin{align}
 Z^{(s)}_B = \prod_{n=s}^{\infty} \frac{1}{|1 - q^n|^2} ~.
 \label{hsb}
\end{align}
In appendix \ref{heat} we show that this
expression actually holds even for $s=1$.

Our theory also includes the fermionic fields \eqref{fermiondeg}.
Before dealing with our theory, let us review the case with
$\text{OSP}(1|2) \otimes \text{OSP}(1|2)$ Chern-Simons theory.
The dual theory should be an ${\cal N}=(1,1)$ superconformal
field theory in two dimensions, and the partition function of the
Chern-Simons theory was proposed in \cite{MW} by making use
of the boundary degrees of freedom. The partition function at one loop
level is
$Z^{(2)}_B Z^{(1)}_F$, where the fermionic part is
\begin{align}
Z^{(1)}_F = \prod_{n=1}^{\infty} |1 + q^{n+\frac12}|^2 ~.
\label{hsf2}
\end{align}
As discussed above, the Chern-Simons theory has
two Majorana gravitini  with the opposite
signature of mass term, and the partition function of the sector was directly
computed in \cite{GGS} as
\begin{align}
Z^{(1)}_F = \frac{\text{det}^{\frac12} \left( - \Delta - \frac{9}{4} \right)^\text{TT}_{(\frac32)}}
{\text{det}^{\frac12} \left( - \Delta + \frac{3}{4} \right)^\text{TT}_{(\frac12)}} ~,
\end{align}
which leads to \eqref{hsf2} with the help of \eqref{hkdet}.

For generic $s \geq 2$, the partition function has not been computed yet.
It will be obtained in the next section, which is
a central result of this note.
As in the bosonic case, we change the basis from
$\psi^{(s)r}_{\mu, \pm }$ to $\psi^{\alpha , \pm}_{\mu a_1 \cdots a_{s-1}}$.
Here the latter is a two component Majorana fermion with $\alpha = 1,2$
and symmetric and traceless for the indices $a_1 , \cdots , a_{s-1}$.
We further define
\begin{align}
 \psi^{\alpha, \pm }_{\mu_1  \cdots  \mu_{s}}
  = \frac{1}{s} \bar e_{(\mu_1}^{~~~a_1} \cdots \bar e_{\mu_{s-1}}^{~~~a_{s-1}}
  \psi^{\alpha, \pm}_{\mu_s ) a_1 \cdots a_{s-1}} ~.
  \label{fermionbasis}
\end{align}
The other fermionic fields
$\tilde \psi^{\alpha, \pm }_{\mu_1  \cdots  \mu_{s}}$ are defined in the
same way. In the next section we find that the one loop determinant is
\begin{align}
  Z^{(s)}_F  =  \frac{\text{det} ^{\frac12} ( - \Delta + (s + \tfrac12) (s - \tfrac52))_{(s+\frac12)}^\text{TT} }{\text{det}  ^{\frac12}( - \Delta + (s - \tfrac12) (s + \tfrac12))_{(s-\frac12)}^\text{TT} } = \prod_{n=s}^{\infty} |1 + q^{n+\frac12}|^2
  \label{hsf}
\end{align}
for a pair of $\psi^{\alpha, a }_{\mu_1  \cdots  \mu_{s}}$ and $\tilde \psi^{\alpha, a }_{\mu_1  \cdots  \mu_{s}}$ with $a=+$ or $a=-$.
Notice that the pair has the opposite signature of mass as mentioned before.

\subsection{Massive scalars and spin $1/2$ spinors}

The Vasiliev's theory includes
massive scalars and fermions coupled to higher spin gauge fields
\cite{PV1,PV2}. Explicitly,
there are 4 complex massive scalars and 4 massive Dirac
fermions  with masses%
\footnote{
According to (3.22) and (3.23) of \cite{PV1}, the masses are
$
 ( M^B_\pm )^2  = \frac{\tilde \lambda^2}{2}\nu ( \nu  \mp 2 ) ,
  (M^F_\pm)^2= \frac{\tilde \lambda^2}{2}\nu^2
$
for bosons and fermions, respectively.
We first need to change the definition of mass for the scalar field to the standard one as
$M^2 - \frac{3}{2} \tilde \lambda^2 \to M^2$ (see (2.19) of \cite{PV1}).
Then we set AdS radius to be one as $(\sqrt{2} \tilde \lambda)^{-1} = 1$ (see below (2.8) of \cite{PV1}).
Finally we change $\frac{1}{2} (\nu+1) = \lambda$.}
\begin{align}
 ( M^B_+ )^2 = - 1 + (\lambda - 1 )^2 ~,
 \qquad
 ( M^B_-)^2 = - 1 + \lambda  ^2 ~,   \qquad
 (M^F_\pm)^2 = (\lambda - \tfrac{1}{2})^2 ~,
\end{align}
where there are two bosons for each of the two masses $M^B_\pm$ and two fermions for each $M^F_\pm$.

As in \cite{GG}, we restrict the range of parameter as $0 \leq \lambda \leq 1$.
For a scalar field with $-1 \leq M^2_+ \leq 0 $ we can choose
two types of boundary conditions.
In the context of the AdS/CFT correspondence, see \cite{KW}.
Since one ${\cal N}=2$ chiral multiplet
includes two scalars and two fermions, we can separate our scalars and fermions
into two groups. Following \cite{GG} we assign the opposite boundary conditions
to the two groups. We have to assign the same type of boundary conditions
to the fields in the same group in order to preserve the ${\cal N}=2$ supersymmetry.
The dictionary between the mass and the conformal dimension of boundary theory
is given by
\begin{align}
(M^B_\pm)^2 = \Delta (\Delta - 2) ~, \qquad
(M^F_\pm)^2 = (\Delta - 1)^2
\end{align}
for a massive scalar and a massive spin $1/2$ fermion, respectively.
The conformal dimensions of dual fields are thus
\begin{align}
 (\Delta^B_+  , \Delta^F_{\pm} , \Delta^B_-) = (2 - \lambda , \tfrac{3}{2} - \lambda , 1 - \lambda )  , ~
 (\lambda , \tfrac{1}{2} + \lambda , 1 + \lambda)
 \label{dualcd}
\end{align}
for each group.

The partition functions for massive scalars and spinors can be computed
by using the heat kernel method as for the higher spin gauge fields.
For a complex scalar field associated with $(h,h)$, the partition function
at one loop
is given in \cite{GMY,GG}
\begin{align}
Z_\text{scalar}^{h} = \prod_{l,l'=0}^\infty \frac{1}{(1 - q^{h+l} \bar q^{h+l'})^2} ~.
\label{hscalar}
\end{align}
For a Dirac spin $1/2$ spinor
associated with $(h, h - 1/2)$ and $(h -1/2 , h )$,
the partition function is computed in
appendix \ref{heat} as
\begin{align}
Z_\text{spinor}^{h } = \prod_{l,l'=0}^\infty ( 1 + q^{h+l} \bar q^{h - \frac12 +l'} )
 ( 1 + q^{h - \frac12 +l} \bar q^{h +l'} ) ~.
 \label{hspinor}
\end{align}

\subsection{A summary of supergravity partition function}

Here we summarize the partition function of the supergravity theory
at the one-loop level.
The contribution from higher spin fields is
\begin{align}
 Z_0   =  \prod_{s=2}^\infty Z^{(s)}_B (Z^{(s-1)}_F)^2 Z^{(s-1)}_B ~,
 \label{zzero}
\end{align}
where the factor from each sector is given in \eqref{hsb} and \eqref{hsf} as
\begin{align}
Z^{(s)}_B = \prod_{n=s}^{\infty} \frac{1}{|1 - q^n|^2} \ ,\qquad
Z^{(s)}_F = \prod_{n=s}^{\infty} |1 + q^{n+\frac12}|^2 \ .
\end{align}
The total contribution is then given by
\begin{align} \label{eq:totalbulk}
 Z^\text{Bulk} &= Z^{\frac{\lambda}{2}}_\text{susy}  Z^{\frac{1-\lambda}{2}}_\text{susy} Z_0 ~,
\end{align}
where the contribution from the massive fields is written as
\begin{align} \label{zsusy}
Z^h_\text{susy}
  = Z_\text{scalar}^{h} ( Z_\text{spinor}^{h + \frac12} )^2 Z_\text{scalar}^{h + \frac12}
\end{align}
with \eqref{hscalar} and \eqref{hspinor}
\begin{align}
Z_\text{scalar}^{h} = \prod_{l,l'=0}^\infty \frac{1}{(1 - q^{h+l} \bar q^{h+l'})^2} \ , \qquad
Z_\text{spinor}^{h } = \prod_{l,l'=0}^\infty ( 1 + q^{h+l} \bar q^{h - \frac12 +l'} )
 ( 1 + q^{h - \frac12 +l} \bar q^{h +l'} ) ~.
\end{align}

{}From the above expression, we can see the relation to the bosonic model
considered in \cite{GG}.
Let us split the partition function of the shs bulk model into the contribution from the bosons and the fermions:
\begin{align}\label{}
    Z^{\textrm{Bulk}}&=Z^{\textrm{Bulk}}_BZ^{\textrm{Bulk}}_F\ ,\\
    Z^{\textrm{Bulk}}_B&=Z_\text{scalar}^{\frac{1 + \lambda}{2}}Z_\text{scalar}^{\frac{ \lambda}{2}}Z_\text{scalar}^{\frac{2 - \lambda}{2}}Z_\text{scalar}^{\frac{1 - \lambda}{2}}Z^{(s=1)}_B\prod_{s=2}^\infty \big(Z^{(s)}_B\big)^2  \ ,\\
    Z^{\textrm{Bulk}}_F&=\big(Z_\text{spinor}^{\frac{1 + \lambda}{2}}\big)^2\big(Z_\text{spinor}^{\frac{2 - \lambda}{2}} \big)^2\prod_{s=1}^\infty \big(Z^{(s)}_F\big)^2  \ .
\end{align}
The bosonic higher spin theory considered by Gaberdiel and Gopakumar with coupling $\lambda$ has one loop partition function \cite{GG}
\begin{align}\label{}
    Z_{\text{GG}}(\lambda)=Z_\text{scalar}^{\frac{1 + \lambda}{2}}Z_\text{scalar}^{\frac{1 - \lambda}{2}}\prod_{s=2}^\infty Z^{(s)}_B\ .
\end{align}
Thus the bosonic part of the partition function is the same as the product of two bosonic higher spin partition functions with parameters $\lambda$ and $1-\lambda$ respectively and an additional spin-1 sector:
\begin{align}\label{eqrelationtoGG}
    Z^{\textrm{Bulk}}_B=Z^{(s=1)}_BZ_{\text{GG}}(\lambda)Z_{\text{GG}}(1-\lambda) ~.
\end{align}

\section{One loop determinant of higher spin fermion}

We compute the partition function for the free theory of fermionic half-integer
spin particles on AdS space as in eq.~\eqref{hsf}.
The partition function for the special case with spin 3/2 gravitino was already obtained
in \cite{DGG}.
For the computation, we follow the strategy of \cite{DGG,GGS}
by making use of the free theory found in \cite{FFdS}.

\subsection{Free theory of higher spin fermion}
\label{freeaction}

First of all, we need the action for the higher spin fermionic fields in the free limit. One possibility is to take the free limit of our Chern-Simons theory.
We choose to take a different route.
At the flat space the free action for higher spin fields can be obtained
uniquely if we assume the gauge symmetry of the form
\begin{align}
\delta \psi_{(s)} = \partial \epsilon_{(s-1)}
\end{align}
as shown in \cite{dWF}.
Then the effects of curvature of AdS space can be
introduced uniquely. Thus we can safely use the action obtained in this way.

We start from the flat space example and then move to AdS space.
We introduce a fully symmetric spinor-tensor $\psi^\alpha_{\mu_1  \cdots \mu_s}$.
Here we use two component Dirac spinors and the spin index will be often suppressed.
The gamma matrices are defined by
$\{\Gamma_\mu , \Gamma_\nu \} = 2 g_{\mu \nu} $.
The field equation is given by \cite{FF} (for a review see \cite{Campoleoni})
\begin{align}
 {\cal S}_{\mu_1 \cdots \mu_s}
  \equiv \slash{\partial} \psi_{\mu_1 \cdots \mu_s}
 - \partial_{ ( \mu_1 } \slashb{\psi} _{\mu_2 \cdots \mu_s )}  = 0 ~.
\end{align}
Here the rule of parentheses is the same as the one adopted in \cite{CFPT}.
Namely, it is a complete symmetrization
of the indices enclosed, with the minimal possible number of terms and without any
normalization.
The equation is invariant under the gauge transformation
\begin{align}
 \delta \psi_{\mu_1 \cdots \mu_s}
  = \partial_{ (\mu_1 } \epsilon_{\mu_2 \cdots \mu_s )} ~, \qquad
  \slash{\epsilon}_{\mu_1 \cdots \mu_{s-2}} = 0 ~. \label{gauge0}
\end{align}
Here the $\Gamma$-traceless condition arises naturally as shown in \cite{dWF}.
If we assign triple $\Gamma$-traceless constraint,
\begin{align}
 { \slashb{\psi}_{\mu_1 \cdots \mu_{s-3} \lambda}  }^{\lambda}
 = \Gamma^{\lambda} \Gamma^{\sigma} \slash \psi_{\mu_1 \cdots \mu_{s-3}
 \lambda \sigma} = 0 ~,
\end{align}
then the Lagrangian \cite{Fronsdal}
\begin{align}
 {\cal L} =  \bar{\psi}^{\mu_1 \cdots \mu_s}
 \left( {\cal S}_{\mu_1 \cdots \mu_s}  - \tfrac12 \Gamma_{(\mu_1} \slash{{\cal S}}_{\mu_2 \cdots \mu_s )} - \tfrac12 \eta_{(\mu_1 \mu_2} {{\cal S}_{\mu_3 \cdots \mu_s ) \lambda} }^{\lambda}
 \right) \label{action0}
\end{align}
leads to the above field equation. The Euclidean signature is assumed for the
space-time, but it is easy to move to the Minkowski space-time.

In AdS space, we should replace the derivative $\partial_\mu$
by the covariant derivative $\nabla_\mu$ on AdS space. A useful formula is
\begin{align}
 [ \nabla_\mu , \nabla_\nu]  \psi_{\mu_1 \cdots \mu_s}
  = \tfrac18 R_{\mu \nu \rho \sigma} [ \Gamma^\rho , \Gamma^\sigma]
     \psi_{\mu_1 \cdots \mu_s} -
  \sum_{i=1}^s  R^\rho_{ ~ \mu_i \mu \nu } \psi_{\mu_1 \cdots \hat \mu_i \cdots \mu_s \rho} ~,
  \label{formula}
\end{align}
where the Riemann curvature tensor for AdS$_3$ is given by
\begin{align}
R_{\mu \nu \rho \sigma}
 = - \tfrac{1}{\ell^2}
 ( g_{\mu \rho} g_{\nu \sigma} - g_{\mu \sigma} g_{\nu \rho} ) ~.
\end{align}
The AdS radius will be set to one, $\ell=1$, as before.
The kinetic term is obtained from \eqref{action0},
or explicitly as
\begin{align}
 {\cal L}_K &= \bar \psi^{\mu_1 \cdots \mu_s} \slash{\nabla} \psi_{\mu_1 \cdots \mu_s}
 - \bar \psi^{\mu_1 \cdots \mu_{s}} \nabla_{( \mu_1} \slashb{\psi}_{\mu_2 \cdots \mu_s)}
 - \bar {\slashb{\psi}}^{ (\mu_1 \cdots \mu_{s-1}}
\nabla^{\mu _s)} \psi_{\mu_1 \cdots \mu_s }  \nonumber \\
  &+ s \bar {\slashb{\psi}}^{\mu_1 \cdots \mu_{s-1}} \slash{\nabla}
  { \slashb \psi}_{\mu_1 \cdots \mu_{s-1}}
  + \tfrac12 \bar \psi^{\mu_1 \cdots \mu_s} \Gamma_{(\mu_1}
  \nabla_{\mu_2} { \psi_{\mu_3 \cdots \mu_{s}) \lambda} }^\lambda  \nonumber \\
 &+ \tfrac12 s ( s-1)  \bar \psi^{\mu_1 \cdots \mu_{s -2 } \lambda}_
{\qquad \quad ~ \lambda}
 \nabla^\sigma \slashb \psi_{\mu_1 \cdots \mu_{s-2}  \sigma}
- \tfrac14 s (s-1) \bar \psi^{\mu_1 \cdots \mu_{s-2} \lambda}_{\qquad \quad ~ \lambda} \slash{\nabla}
 { \psi_{\mu_1 \cdots \mu_{s-2} \sigma}}^{\sigma} ~.
\end{align}
The naive gauge transformation may be given by \eqref{gauge0}
with the derivative replaced by the covariant derivative.
However, the above kinetic term is not invariant under the
 transformation since the covariant derivative
does not commute with each other.
In order to make the action gauge invariant, we add a mass term as
\begin{align}
{\cal L} = {\cal L}_K + {\cal L}_M \label{action} ~,
\end{align}
where
\begin{align}
\label{massterm}
{\cal L}_M = \zeta \tfrac{2s - 1}{2}  \left(
 \bar \psi_{\mu_1 \cdots \mu_s} \psi^{\mu_1 \cdots \mu_s} - s
\bar {\slashb{\psi}}^{\mu_1 \cdots \mu_{s-1}}
\slashb{\psi}_{\mu_1 \cdots \mu_{s-1}}
-  \tfrac14 s (s-1) { \bar \psi^{\mu_1 \cdots \mu_{s-2} \lambda}}_{\qquad \quad ~ \lambda}
 { \psi_{\mu_1 \cdots \mu_{s-2} \sigma}}^{\sigma} \right) ~.
\end{align}
Moreover, we shift the gauge transformation as
\begin{align}
 \delta \psi_{\mu_1 \cdots \mu_s}
 = \nabla_{( \mu_1} \epsilon_{\mu_2 \cdots \mu_s )} + \zeta \tfrac12  \Gamma_{( \mu_1}
\epsilon_{ \mu_2 \cdots \mu_s )} ~.
\label{dgt}
\end{align}
Here we can take the both signs of mass as $\zeta = \pm$.
The conditions
\begin{align}
 { \slashb{\psi}_{\mu_1 \cdots \mu_{s-3} \lambda}  }^{\lambda} = 0 ~,
\qquad \slash{\epsilon}_{\mu_1 \cdots \mu_{s-2}} = 0 \label{}
\label{triplegamma}
\end{align}
are the same as in the flat space case.

Before moving to the analysis of the free theory, let us remark the
relation to the Chern-Simons formulation.
First, the triple $\Gamma$-traceless
constraint \eqref{triplegamma} can be obtained from
the traceless condition for $a_i$ indices of
$\psi_{\mu a_1 \cdots a_{s-1}}$ through the change of basis \eqref{fermionbasis}.
Next, the relation of fields can be obtained as follows.
In a Euclidean space, we may treat $\psi_{(s)}$ and
$\bar \psi_{(s)}$ as independent variables. Integrating by parts, we can see
that the Dirac equation for each field has the opposite signature of the mass term.
Thus two Majorana fermions with opposite signature of mass in the
Chern-Simons formulation should be identified with
one Dirac fermion in the theory obtained in this subsection.
In order to relate the equations of motion of both theories,
we have to define dual variables
such as ${\psi}^{'}_\mu = \epsilon_{\mu \nu \rho} \Gamma^\nu \psi^{\rho}$.

\subsection{Degrees of freedom}

Let us count the degrees of freedom as in \cite{GGS}.
A complete symmetric tensor-spinor of rank $s$ in three dimensions has
$(s+1)(s+2)/2 \times 2$ components. The multiplication of $2$ arises since
 we are dealing with
two component spinors. On the other hand, there are as many triple $\Gamma$-traceless
constraints
as rank $(s-3)$ tensor-spinor components.
Therefore, the total number of components is $6s$. The gauge parameter is
a complete symmetric tensor-spinor of rank $(s-1)$ subject to
$\Gamma$-traceless constraints. Thus the number of independent gauge parameters
is $2s$. Now the action for the spinor is of first order,
so there are $2s$ constraints \cite{dWF}. In this way, we have a topological
theory without any propagating modes.

As emphasized in \cite{GGS}, it is an crucial task to decompose the
fields in an appropriate way.
Following the analysis of the gravitino in \cite{DGG},
we decompose the fields as
\begin{align}
 \psi_{\mu_1 \cdots \mu_s} = \Gamma_{( \mu_1} \hat
 \psi_{\mu_2 \cdots \mu_s )} + \psi^\text{T} _{\mu_1 \cdots \mu_s} ~,
 \qquad \slashb \psi^\text{T} _{\mu_1 \cdots \mu_{s-1}} = 0 ~.
 \label{dec0}
\end{align}
We decompose furthermore the $\Gamma$-traceless part as
\begin{align}
 \psi^\text{T}_{\mu_1 \cdots \mu_s}
 = \psi^\text{TT}_{\mu_1 \cdots \mu_s} +
 \psi^{(\eta)}_{\mu_1 \cdots \mu_s } ~,
 \label{dec}
\end{align}
and
\begin{align}
\psi^{(\eta)}_{\mu_1 \cdots \mu_s }
= \nabla_{(\mu_1} \eta_{\mu_2 \cdots \mu_s )}
- \tfrac{1}{2s+1} \left[
 \Gamma_{(\mu_1} \slash \nabla \eta_{\mu_2 \cdots \mu_s )}
 + 2 \nabla^\lambda g_{(\mu_1 \mu_2} \eta_{\mu_3 \cdots \mu_s ) \lambda}
\right]~.
\label{long}
\end{align}
Here  $\eta_{(s-1)}$ satisfies the $\Gamma$-traceless condition as
\begin{align}
\slash \eta_{\mu_1 \cdots \mu_{s-2}} = 0 ~.
\end{align}
We see that $2$ components are transverse $\Gamma$-traceless as
\begin{align}
 \nabla^\lambda \psi^\text{TT}_{\mu_1 \cdots \mu_{s-1} \lambda}
 =  \Gamma^\lambda \psi^\text{TT}_{\mu_1 \cdots \mu_{s-1} \lambda}  = 0 ~,
\end{align}
and $2s$ components of $\eta_{(s-1)}$ are longitudinal and $\Gamma$-traceless.
The other $(4s - 2)$ components are in the $\Gamma$-trace part and the triple
$\Gamma$-traceless condition implies that
\begin{align}
  { \hat \psi}_{\mu_1 \cdots \mu_{s-3} \lambda} ^{ \qquad \quad ~ \lambda} = 0 ~.
\end{align}

\subsection{One loop determinant}

In the path integral formulation, we can obtain the partition function
as the product of one loop determinants by integrating over the fields.
We compute the partition function by separating the field into three parts.
As in \eqref{dec0} we decompose the field into $\Gamma$-trace part
$\hat \psi_{(s-1)}$ and $\Gamma$-traceless part  $\psi^\text{T}_{(s)}$.
Moreover, we have to fix the gauge, then the contribution from the
Faddeev-Popov ghosts arises.
We first consider the $\Gamma$-traceless part.
After fixing the gauge, we compute the ghost contribution,
then we evaluate the $\Gamma$-trace part.
Finally we combine these three parts.

\subsubsection{The $\Gamma$-traceless part}

We start from the $\Gamma$-traceless part with $\psi^\text{T}_{(s)}$ in \eqref{dec0}.
The Lagrangian \eqref{action} for this part is simplified as
\begin{align}
 {\cal L} = {\bar \psi}^{ \text{T} \mu_1 \cdots \mu_s}
 ( \slash \nabla + \zeta (s - \tfrac12 ) ) \psi^\text{T} _{\mu_1 \cdots \mu_s} ~.
 \label{actiontl}
\end{align}
The  $\Gamma$-traceless part is also divided into the transverse modes and
the longitudinal modes as in \eqref{dec}.
For the transverse traceless components,
we have the one loop determinant
\begin{align}
Z(\psi^\text{TT}_{(s)}) = \text{det} (\slash \nabla + \zeta ( s - \tfrac12 ) )^\text{TT}_{(s+\frac12)} ~ .
\end{align}
Here the subscripts imply that the derivative acts on transverse traceless part of spin $(s+1/2)$ spinor.

For the longitudinal modes, we should put the expression of \eqref{long}
into the action \eqref{actiontl}.
In order to do so, we may compute
\begin{align}
\label{Deta}
  ( {\cal D}^{(\eta)} \psi^{(\eta)} )_{\mu_1 \cdots \mu_s}
 &= ( \slash \nabla + \zeta ( s - \tfrac12 )) \psi^{(\eta)} _{\mu_1 \cdots \mu_s}
 \\ \nonumber
  &\simeq \left( \frac{s - \frac12}{s + \frac12} \right)
 \nabla_{(\mu_1} (\slash \nabla + \zeta (s + \tfrac12 ) ) \eta_{\mu_2 \cdots \mu_s )}~,
\end{align}
where $\simeq$ means equality up to $\Gamma$-exact term.
The action is obtained by multiplying $\bar \psi^{(\eta)}_{(s)}$
from the left, so the $\Gamma$-exact term is irrelevant.
Here we have changed the modes from $\psi^{(\eta)}_{(s)}$ to $\eta_{(s)}$,
thus we have to take into account the Jacobian $Z^{(\eta)}_{(s)}$.
One way to get the Jacobian is to compute
\begin{align}
 1 = \int [ {\cal D}  \psi^{(\eta)}_{(s)} ]
 e^{- \langle \psi^{(\eta)}_{(s)} , \psi^{(\eta)}_{(s)} \rangle}
   = Z^{(\eta)}_{(s)} \int  [ {\cal D}  \eta]
 e^{- \langle \psi^{(\eta)}_{(s)} , \psi^{(\eta)}_{(s)} \rangle} ~.
\end{align}
Now the Gaussian weight is
\begin{align}
 & - \langle \psi^{(\eta)}_{(s)} , \psi^{(\eta)}_{(s)} \rangle
    = - \bar \psi^{(\eta) \mu_1 \cdots \mu_s} \nabla_{( \mu_1 } \eta_{  \mu_2 \cdots \mu_{s} )}
\\
 &  =  \left( \bar \eta^{ ( \mu_1 \cdots \mu_{s-1} } \nabla^{\mu_s )}
- \tfrac{1}{2s+1} \left[ \bar \eta^{ ( \mu_1 \cdots \mu_{s - 1}} \slash \nabla
 \Gamma^{\mu_s ) }
 + 2 \bar \eta^{ \lambda( \mu_1 \cdots \mu_{s-2}  } g^{\mu_{s-1} \mu_s )} \nabla_\lambda \right] \right)
 \nabla_{( \mu_1 } \eta_{  \mu_2 \cdots \mu_{s} )} ~.
 \nonumber
\end{align}
{}From this we can see that the complicated contribution from the multiplication
of $\bar \psi^{(\eta)}_{(s)}$ to \eqref{Deta} cancels the Jacobian.
Therefore, the contribution from the longitudinal modes is summarized as
\begin{align}
Z (\eta_{(s-1)}) = \text{det} (\slash \nabla + \zeta ( s + \tfrac12 ) )^\text{T}_{(s-\frac12)} ~,
 \label{etacont}
\end{align}
where the subscripts imply that the derivative acts on traceless part of spin $(s-1/2)$ spinors.

\subsubsection{Ghost contribution}

If we naively put the $\Gamma$-trace part into the action \eqref{action}, then
we would have a quite complicated expression.
In order to simplify the computation, we choose a convenient gauge.
With the decomposition of \eqref{dec0} and \eqref{dec},
the gauge transformation is given by
\begin{align}
  \delta \hat \psi_{\mu_1 \cdots \mu_{s-1}} = \tfrac{1}{2s+1}
  \left[ (\slash \nabla + \zeta ( s + \tfrac12 ) ) \epsilon_{\mu_1 \cdots \mu_{s-1}}
 + \Gamma_{( \mu_1} \nabla^\lambda \epsilon_{\mu_2 \cdots \mu_{s-1} ) \lambda}  \right] \label{gaugetf}
\end{align}
in addition to
\begin{align}
 \delta \psi^\text{TT}_{\mu_1 \cdots \mu_s}  = 0 ~, \qquad
 \delta \eta_{\mu_1 \cdots \mu_{s-1}}  = \epsilon_{\mu_1 \cdots \mu_{s-1}}  ~.
\end{align}
With this gauge transformation, we set the form of $\hat \psi_{(s-1)}$ as
\begin{align}
 \hat \psi_{\mu_1 \cdots \mu_{s-1}} = \tfrac12
\Gamma_{( \mu_1} \tilde \psi_{\mu_2 \cdots \mu_{s-1} ) } ~ ,\qquad \tilde { \slashb \psi } _{\mu_1 \cdots \mu_{s-3}} = 0 ~.
\label{gaugefix}
\end{align}
{}From \eqref{gaugetf} the Faddeev-Popov determinant is found to be
\begin{align}
Z_\text{gh} = \text{det}^{-2} (\slash \nabla + \zeta ( s + \tfrac12) )_{(s-\frac12)}^\text{T} ~.
\end{align}
Combining the contribution from $\eta_{(s-1)}$ \eqref{etacont}, we have
\begin{align}
Z_\text{gh} Z (\eta_{(s-1)})
= Z_\text{gh}^{\frac12} = \text{det}^{-1} (\slash \nabla + \zeta ( s + \tfrac12 ) )_{(s-\frac12)}^\text{T} ~.
\end{align}
Since the formula \eqref{hkdet} can be applied only for transverse modes, we should
rewrite this determinant furthermore.

Introduce a Dirac tensor-spinor $\theta_{(s-1)}$ of rank $(s-1)$
subject to the $\Gamma$-traceless constraint
\begin{align}
 \slash \theta_{\mu_1 \cdots \mu_{s-2}} = 0 ~.
\end{align}
Then the one loop determinant of the ghost can be written as
\begin{align}
 1 = Z_\text{gh}^{\frac12} \int [{\cal D} \theta_{(s-1)}] e^{- \int d ^3 x {\cal L}^\theta} ~,
 \qquad {\cal L}^\theta = \bar \theta^{\mu_1 \cdots \mu_{s-1}}
 (\slash \nabla + \zeta ( s + \tfrac12 ) )  \theta_{\mu_1 \cdots \mu_{s-1}} ~.
\end{align}
As always, we decompose $\theta_{(s-1)}$ into the transverse modes and the
longitudinal modes as
\begin{align}
 \theta_{\mu_1 \cdots \mu_{s-1}}
 = \theta^\text{TT} _{\mu_1 \cdots \mu_{s-1}} + \theta^{(\xi)}_{\mu_1 \cdots \mu_{s-1}} ~,
 \qquad \nabla^\lambda \theta^\text{TT} _{\mu_1 \cdots \mu_{s-2} \lambda} = 0 ~.
\end{align}
Here the $\Gamma$-traceless condition leads to
\begin{align}
\theta^{(\xi)}_{\mu_1 \cdots \mu_{s-1} }
= \nabla_{(\mu_1} \xi_{\mu_2 \cdots \mu_{s-1} )}
- \tfrac{1}{2s-1} \left[
 \Gamma_{(\mu_1} \slash \nabla \xi_{\mu_2 \cdots \mu_{s-1} )}
 + 2 \nabla^\lambda g_{(\mu_1 \mu_2} \xi_{\mu_3 \cdots \mu_{s-1} ) \lambda}
\right]~.
\end{align}
For the transverse part $\theta^\text{TT}_{(s-1)}$, the one loop contribution is
\begin{align}
 \text{det} (\slash \nabla + \zeta ( s + \tfrac12 ) )_{(s-1)}^\text{TT} ~.
\end{align}
Precisely speaking, we are interested in the ghost contribution, which is
the inverse of it.
For the longitudinal part $\theta^{(\xi)}_{(s-1)}$, we compute
\begin{align}
 ( {\cal D}^{(\xi)} \theta^{(\xi)} )_{\mu_1 \cdots \mu_{s-1}}
 &= ( \slash \nabla + \zeta ( s + \tfrac12 ) ) {\theta^{(\xi)} }_{\mu_1 \cdots \mu_{s-1}}  \\
 & \simeq \left( \frac{s - \frac32}{s - \frac12} \right)
 \nabla_{(\mu_1} \left(\slash \nabla + \zeta \frac{s^2 - \tfrac14}{s - \frac32} \right) \xi_{\mu_2 \cdots \mu_{s-1} )}  \nonumber
\end{align}
as for $\eta_{(s-1)}$. The Jacobian due to the change from $\theta^{(\xi)}_{(s-1)}$
to $\xi_{(s-2)}$ can be analyzed in the
same way as for $\eta_{(s-1)}$, and the one loop contribution from this term
is found as
\begin{align}
 \text{det} \left(\slash \nabla + \zeta \frac{s^2 - \tfrac14}{s - \frac32} \right)_{(s-\frac32)}^\text{T} ~,
\end{align}
or its inverse for the corresponding ghost contribution.
Totally, we may rewrite the contribution as
\begin{align}
Z_\text{gh}^{\frac12} =  \text{det} ^{-1} (\slash \nabla + \zeta ( s + \tfrac12 ) )_{(s-\frac12)}^\text{TT}
 \text{det} ^{-1} \left(\slash \nabla + \zeta \frac{s^2 - \tfrac14}{s - \frac32} \right)_{(s-\frac32)}^\text{T} ~.
\end{align}
Next we will see that the contribution from the $\Gamma$-trace modes cancels
the one from $\theta^{(\xi)}_{(s-1)}$.

\subsubsection{The $\Gamma$-trace part}

We fixed the gauge as in \eqref{gaugefix} such that the expression becomes simpler.
Let us see that this is indeed the case.
Due to the gauge fixing,
we just need to compute the action in terms of
\begin{align}
 \psi_{\mu_1 \cdots \mu_s} = g_{(\mu_1 \mu_2} \tilde \psi_{\mu_3 \cdots \mu_s )} ~.
\end{align}
We start from the kinetic term.
It is useful to use the expression of \eqref{action0} with the covariant derivative,
since now we have
\begin{align}
 {{\cal S}_{\mu_1 \cdots \mu_{s-2} \lambda} }^\lambda
 &= \slash \nabla { \psi_{\mu_1 \cdots \mu_{s-2} \lambda} } ^ \lambda - 2 \nabla^\lambda
  \slashb \psi_{\mu_1 \cdots \mu_{s-2} \lambda} \\
 &= (2 s - 3) \slash \nabla \tilde \psi_{\mu_1 \cdots \mu_{s-2}}
 - 2 \Gamma_{( \mu_1} \nabla^\lambda \tilde \psi_{\mu_2 \cdots \mu_{s-2} ) \lambda} ~.
 \nonumber
\end{align}
With the new variable the kinetic term becomes
\begin{align}
 {\cal L}_K &= - \frac{s(s-1)(2s-1)}{4} \bar {\tilde \psi }^{\mu_1 \cdots \mu_{s-2}}
 {{\cal S}_{\mu_1 \cdots \mu_{s-2} \lambda} }^\lambda \\
 &= - \frac{s(s-1)(2s-1)(2s-3)}{4} \bar {\tilde \psi }^{\mu_1 \cdots \mu_{s-2}}
  \slash \nabla
 \tilde \psi_{\mu_1 \cdots \mu_{s-2}}  ~. \nonumber
\end{align}
Next we move to the mass term.
The expression of \eqref{massterm} reduces to
\begin{align}
{\cal L}_M &= \zeta \frac{2s - 1}{2}
 \bar \psi^{\mu_1 \cdots \mu_s}
\left(
 \psi_{\mu_1 \cdots \mu_s} - \Gamma_{(\mu_1 }
 \slashb \psi_{\mu_2 \cdots \mu_{s})}
-  \tfrac12 g_{(\mu_1 \mu_2} { \psi_{\mu_3 \cdots \mu_{s} ) \sigma}}^{\sigma} \right) \\
&= -  \zeta  \frac{(2s - 1) s (s-1) (2s+1)}{4}
 \bar { \tilde \psi}^{\mu_1 \cdots \mu_{s-2}} {\psi_{\mu_1 \cdots \mu_{s-2} \lambda}}^\lambda
\nonumber \\
 &= - \zeta  \frac{(2s - 1)^2 s (s-1) (2s+1) }{4}
 \bar { \tilde \psi}^{\mu_1 \cdots \mu_{s-2}} { \tilde \psi}_{\mu_1 \cdots \mu_{s-2}} ~.
 \nonumber
\end{align}
Thus, the action is quite simple in the gauge fixing \eqref{gaugefix} as expected.
The one loop contribution from this part is obtained as
\begin{align}
Z(\tilde \psi_{(s-2)}) = \text{det}  \left(\slash \nabla + \zeta \frac{s^2 - \tfrac14}{s - \frac32} \right)_{(s-\frac32)}^\text{T} ~.
\end{align}
This cancels the contribution from $\theta^{(\xi)}_{(s-1)}$ as mentioned above.

\subsubsection{Total contribution}

Our Chern-Simons theory has four Majorana tensor-spinors with spin $s+1/2$,
and the mass of two spinors has opposite signature compared to
those of the other two.
{}From the argument in the end of section \ref{freeaction}, they are mapped
to two Dirac tensor-spinors of rank $s$ in the free theory \eqref{action}.
For one Dirac fermion, the partition function is
obtained as
\begin{align}
  Z^{(s)}_F
 = \frac{\text{det} ( \slash \nabla + \zeta (s - \tfrac12))^\text{TT}_{(s+\frac12)}}
  {\text{det} ( \slash \nabla + \zeta (s + \tfrac12))^\text{TT}_{(s-\frac12)}}
\end{align}
by combining the results obtained so far.
As in appendix D of \cite{DGG}, we have
\begin{align}
 - (\slash \nabla + \hat m) (\slash \nabla - \hat m) \psi^\text{TT} _{\mu_1 \cdots \mu_s}
 = (-  \Delta - s - \tfrac32 + \hat m^2) \psi^\text{TT} _{\mu_1 \cdots \mu_s}~,
 \label{shiftmass}
\end{align}
where we define $\Delta = \nabla^\mu \nabla_\mu$.
Thus the one loop contribution is written as
\begin{align}
  Z^{(s)}_F  =  \frac{\text{det}^{\frac12} (- \Delta + (s + \tfrac12) (s - \tfrac52))_{(s+\frac12)}^\text{TT} }{\text{det}^{\frac12} (- \Delta + (s - \tfrac12) (s + \tfrac12))_{(s-\frac12)}^\text{TT} } ~.
\end{align}
The application of the heat kernel method in appendix \ref{heat} leads to
\begin{align}
\log Z^{(s)}_F &= - \frac12 \sum_{m=1}^\infty \frac{(-1)^m }{m|\sin \frac{m \tau}{2}|^2}
 \left[ \cos ((s+\tfrac12) m \tau_1)    e^{- m \tau_2 (s - \tfrac12)}
 - \cos ((s-\tfrac12) m \tau_1)    e^{- m \tau_2 (s + \tfrac12)} \right]
 \nonumber \\
  &= - \sum_{m=1}^\infty \frac{(-1)^m}{m} \left[ \frac{q^{m(s+\frac12)}}{1 - q^m}
 + \frac{\bar q^{m(s+\frac12)}}{1 - \bar q^m} \right]
 = \log \left[ \prod_{n=s}^\infty | 1 + q^{n+\frac12} |^2 \right] ~.
\end{align}
Therefore, we have
\begin{align}
Z^{(s)}_F
 = \prod_{n=s}^\infty | 1 + q^{n+\frac12} |^2
 \label{sfermion}
\end{align}
as stated in \eqref{hsf}.

\section{Dual ${\cal N }=2$ $\mathbb C$P$^N$ model }

As dual CFT we propose the ${\cal N }=2$ $\mathbb C$P$^N$ model, that is the Kazama-Suzuki coset \cite{KS} (for a review see, e.g.,  \cite{Nozaki})
\begin{align}\label{eq:KScoset}
 \frac{\text{SU}(N+1)_k \times \text{SO}(2 N)_1 }{\text{SU}(N)_{k+1} \times \text{U}(1)_{N(N+1)(k+N+1)} } \, .
\end{align}
The central charge of this model is
$c = 3 N k /(k+N+1)$.
We are interested in the 't Hooft limit
$N,k \to \infty$ keeping
\begin{align}
 \lambda = \frac{N}{k+N}
\end{align}
finite. Note that the central charge is invariant under the exchange of the level $k$ and the rank $N$, we will comment on this level-rank duality below. Keeping this symmetry explicit, the central charge scales as
\begin{equation}
c \, \sim \, 3(1-\lambda)N \, \sim \, 3\lambda(1-\lambda)(N+k)\, .
\end{equation}

\subsection{The vacuum character}

Ito \cite{Ito} showed that the symmetry of the $\mathcal N=2$ $\mathbb C$P$^N$ coset is the $\mathcal N=2$ ${\cal W}_{N+1}$ algebra.
This algebra is defined as the Drinfeld-Sokolov reduction of $\mathfrak g=\text{sl}(N+1|N)$ corresponding to
the principal embedding of sl(2).
Denote by $\underline m$ the $m$ dimensional representation of sl(2). Then the adjoint of $\mathfrak g$ decomposes as
\begin{equation}
\mathfrak g_0 \ = \ \underline 1 \oplus \underline{2N+1}\oplus 2\times\bigoplus_{s=1}^{N-1} \underline{2s+1} ~, \qquad
\mathfrak g_1 \ = \ 2\times\bigoplus_{s=1}^{N} \underline{2s} ~.
\end{equation}
Here $\mathfrak g_0$ and $\mathfrak g_1$ are Grassmann even and odd, respectively. See \eqref{slexp} as well.
Thus the $\mathcal N=2$ super ${\cal W}_{N+1}$ algebra is generated by bosonic fields, two for each spin from 2 to $N$, and one of spin 1 and of spin $N+1$.
These generate the bosonic subalgebra $\hat{\mathfrak u}(1)\oplus {\cal W}_N\oplus {\cal W}_{N+1}$.
The fermionic fields come in pairs of spin from $3/2$ to $N+1/2$.
Using Poincare-Birkhoff-Witt basis, the generic vacuum character is
\begin{equation}
\chi_0 \ = \ \prod_{m=1}^\infty \frac{1}{|1-q^m|^2}\prod_{s=2}^{N}\prod_{m=s}^\infty \frac{1}{|1-q^m|^2} \prod_{s=2}^{N+1}\prod_{m=s}^\infty \frac{1}{|1-q^m|^2}
\prod_{s=1}^{N}\prod_{m=s}^\infty |1+q^{m+\frac12}|^4 ~.
\end{equation}
As in \cite{GG}, we assume that the null vectors do not modify the answer in the 't Hooft limit.
Comparing with the bulk, we see
\begin{equation}
\lim_{N\rightarrow\infty}\chi_0 \ = \ Z^{(s=1)}_B\prod_{s=2}^\infty \big(Z^{(s)}_B\big)^2\prod_{s=1}^\infty \big(Z^{(s)}_F\big)^2 \ =\  Z_0\, .
\end{equation}
where $Z_0$ is defined in \eqref{zzero}.
This is in perfect agreement with the massless part of the bulk higher spin theory.

\subsection{Supersymmetry considerations of the partition function}

Let us recall that the bulk partition function is written
in \eqref{eq:totalbulk} in the following way
\begin{equation}\label{eqsusypartition}
 Z^{\textrm{Bulk}} \ = \ Z_{\text{susy}}^{\frac{\lambda}{2}}Z_{\text{susy}}^{\frac{1-\lambda}{2}} Z_0
\end{equation}
with \eqref{zsusy}
\begin{equation}
\begin{split}
Z_{\text{susy}}^{h} \ &= \ \prod_{l,l'=0}^\infty \frac{( 1 + q^{h+\frac{1}{2}+l} \bar q^{h +l'} )^2
 ( 1 + q^{h +l} \bar q^{h +\frac{1}{2}+l'} )^2}{(1 - q^{h+l} \bar q^{h+l'})^2(1 - q^{h+\frac{1}{2}+l} \bar q^{h+\frac{1}{2}+l'})^2} \\
&= \ Z_{\text{scalar}}^{h}Z_{\text{scalar}}^{h+\frac{1}{2}}\bigl(Z_{\text{spinor}}^{h+\frac{1}{2}}\bigr)^2\, .
\end{split}
\end{equation}
Here $Z_{\text{susy}}^{h}$ shows signatures of supersymmetry.
If one assumes that the states in $Z_{\text{scalar}}^{h}$ are the bottom components of $\mathcal N=(1,1)$ multiplets,
then acting with the left- and right-supercharges one obtains the remaining parts of the partition function.

We are considering the $\mathcal N=(2,2)$ $\mathbb C$P$^N$ sigma model.
The bosonic $\mathbb C$P$^N$ sigma model has the following level-rank duality \cite{Gepner:1988wi}
\begin{align}\label{eq:tduality}
    \frac{\text{SU}(N+1)_k}{\text{SU}(N)_{k} \times \text{U}(1)_{N(N+1)k}}\simeq \frac{\text{SU}(k)_{N} \times \text{SU}(k)_{1} }{\text{SU}(k)_{N+1} } ~.
\end{align}
Note, that the world-sheet supersymmetric sigma model of the level-rank dual coset does not possess extended $\mathcal N=(2,2)$ superconformal symmetry,
but only $\mathcal N=(1,1)$ superconformal symmetry.
This is a well-known issue for level-rank dual theories \cite{Altschuler:1988mg,Bowcock:1988vs}.
The level-rank duality is merely a one-to-one correspondence of the branching functions i.e. of structure of the spectrum. Now the level-rank dual bosonic coset model is precisely the coset model models considered in \cite{GG}. Due to the exchange of the level and the rank the 't Hooft parameter takes the form
\begin{align}\label{}
    \lambda_2=\lim\frac{k}{N+k}=1-\lambda\,.
\end{align}
We know, that the partition function takes the form \cite{GG}
\begin{align}\label{eqGGdual}
    Z_{\text{GG}}(1-\lambda)=Z_\text{scalar}^{h=\frac{2 - \lambda}{2}}Z_\text{scalar}^{h=\frac{\lambda}{2}}\prod_{s=2}^\infty Z^{(s)}_B\ .
\end{align}
The last part is exactly the vacuum character of the $W_\infty[1-\lambda]$-algebra.
The first two parts correspond to the scalars in the bulk theory.
Let us now look back to our supersymmetric partition function \eqref{eqsusypartition}. We see that the first part corresponds to the top
components of the $\mathcal N=(1,1)$ supersymmetry multiplet of weight $h=(1-\lambda)/2$, while
the second
part corresponds to the bottom
components of the $\mathcal N=(1,1)$ supersymmetry multiplet of weight $h=\lambda/2$.

\subsubsection{Decomposing coset characters}

Following \cite{DiFrancesco:1993dg} we use the conformal embedding of $\text{SU}(N)_1\times U(1)_N\rightarrow\text{SO}(2N)_1$. In the action this is simply the rewriting of the Majorana fermions into Dirac fermions. Note that the $\text{SU}(N)$ and $\text{U}(1)$ in the denominator is naturally embedded separately into the two factors. If we now decompose the SU$(N+1)_k$ according to $\text{SU}(N)_k\times \text{U}(1)_{N(N+1)k}$, then the denominator factors are also naturally embedded  into the two factors  separately. This gives the following factorization
\begin{align}\label{eq:cosetdec}
    \frac{\text{SU}(N+1)_k \times \text{SO}(2 N)_1 }{\text{SU}(N)_{k+1} \times \text{U}(1)_{N(N+1)(k+N+1)} }\sim         \frac{\text{SU}(N+1)_k}{\text{SU}(N)_{k} \times \text{U}(1)_{N(N+1)k}}\times \frac{\text{SU}(N)_{k} \times \text{SU}(N)_{1} }{\text{SU}(N)_{k+1} }\times \text{U}(1)
\end{align}
which is just a shorthand for the glueing of characters of the coset.
For the first coset we do a rank-level duality or T-equivalence as in last subsection
\begin{align}\label{}
    \frac{\text{SU}(N+1)_k}{\text{SU}(N)_{k} \times \text{U}(1)_{N(N+1)k}}\simeq \frac{\text{SU}(k)_{N} \times \text{SU}(k)_{1} }{\text{SU}(k)_{N+1} }
\end{align}
and we obtain
\begin{align}\label{eq:cosetdec2}
    \frac{\text{SU}(N+1)_k \times \text{SO}(2 N)_1 }{\text{SU}(N)_{k+1} \times \text{U}(1)_{N(N+1)(k+N+1)} }\simeq           \frac{\text{SU}(k)_{N} \times \text{SU}(k)_{1} }{\text{SU}(k)_{N+1} }\times \frac{\text{SU}(N)_{k} \times \text{SU}(N)_{1} }{\text{SU}(N)_{k+1} }\times \text{U}(1)\, .
\end{align}
Note that in this case we have just used, the level-rank duality is a one-to-one correspondence of branching functions.
To our knowledge the exact form of this correspondence is not known.
However, this clearly means, that the character in the 't Hooft limit of the $\mathcal N=(2,2)$ $\mathbb C$P$^N$ sigma model
can be glued together from a U(1)-character and characters of the $W_\infty[1-\lambda]$ and $W_\infty[\lambda]$.
These characters were calculated in \cite{GGHR}.
Looking back at \eqref{eqrelationtoGG}, we see that the partition functions of these two models give the bosonic part of the bulk partition function
(up to the U(1) vacuum character) of our higher spin supergravity theory.

Alternatively one can start by considering another level-rank duality. According to \cite{Naculich:1997ic} there is strong evidence that the super versions of the complex Grassmanians
\begin{align}
 G(m,n,l)=\frac{\text{SU}(m+n)_l \times \text{SO}(2 mn)_1 }{\text{SU}(m)_{n+l} \times\text{SU}(n)_{m+l}\times \text{U}(1)_{mn(m+n)(m+n+l)} }
\end{align}
with central charges
\begin{align}
 c(m,n,l)=\frac{3mnl}{m+n+l}
\end{align}
have duality in the permutation of $m,n,l$. Besides the obvious permutation of $m,n$ this gives dualities
\begin{align}\label{eq:dualities}
 G(m,n,l)\simeq G(m,l,n)\simeq G(l,n,m) ~.
\end{align}
Our coset \eqref{eq:KScoset} can be written as $G(1,N,k)$. The first relation is thus the level rank duality mentioned in the beginning of the section which exchanges $N$ and $k$.
\begin{align}
  \frac{\text{SU}(N+1)_k \times \text{SO}(2 N)_1 }{\text{SU}(N)_{k+1} \times \text{U}(1)_{N(N+1)(k+N+1)} } &= G(1,N,k) \simeq G(1,k,N) \nonumber \\
&=\frac{\text{SU}(k+1)_N \times \text{SO}(2 k)_1 }{\text{SU}(k)_{N+1} \times \text{U}(1)_{k(k+1)(k+N+1)} } ~.
\end{align}
This suggests that our theory in the 't Hooft limit is invariant under the exchange
\begin{align}
 \lambda\leftrightarrow 1-\lambda ~.
\end{align}
Indeed, we get a strong confirmation of this from the bulk side of the theory where the partition function has this invariance \eqref{eq:totalbulk}.

The second duality in \eqref{eq:dualities} gives us the following relation
\begin{align}
  \frac{\text{SU}(N+1)_k \times \text{SO}(2 N)_1 }{\text{SU}(N)_{k+1} \times \text{U}(1)_{N(N+1)(k+N+1)} } &= G(1,N,k) \simeq G(k,N,1) \\
\label{eq:2ndcoset} &=\frac{\text{SU}(N+k)_1 \times \text{SO}(2 N k)_1 }{\text{SU}(k)_{N+1} \times\text{SU}(N)_{k+1}\times \text{U}(1)_{Nk(N+k)(N+k+1)} } ~.
\nonumber
\end{align}

One advantage in considering this coset as our starting point is that if we want to calculate some branching functions, we only have to look at embedding into level one algebras.
Further, we can now  rewrite the characters of the coset using the following three conformal embeddings (see e.g. \cite{Altschuler:1988mg}):
\begin{align}
 \text{SU}(Nk)_1\times \text{U}(1)_{Nk}&\mapsto\text{SO}(2Nk)_1\, ,\nonumber \\
 \text{SU}(N)_k\times\text{SU}(k)_N&\mapsto\text{SU}(Nk)_1\, ,\nonumber \\
 \text{SU}(N)_1\times\text{SU}(k)_1\times \text{U}(1)_{Nk(N+k)} &\mapsto\text{SU}(N+k)_1 \, .
\end{align}
The glueing of the characters then takes the same form as above
\begin{multline}\label{eq:glue2}
\frac{\text{SU}(N+1)_k \times \text{SO}(2 N)_1 }{\text{SU}(N)_{k+1} \times \text{U}(1)_{N(N+1)(k+N+1)} }    \\
\sim \frac{\text{SU}(k)_{N} \times \text{SU}(k)_{1} }{\text{SU}(k)_{N+1} } \times\frac{\text{SU}(N)_{k} \times \text{SU}(N)_{1} }{\text{SU}(N)_{k+1} }\times \frac{\text{U}(1)_{Nk(N+k)}\times \text{U}(1)_{Nk}}{\text{U}(1)_{Nk(N+k)(N+k+1)}} \, ,
\end{multline}
where the U(1) in the denominator has level $Nk(N+k)$ in the direction of the first U(1) and level $Nk(N+k)^2$ in the direction of the second U(1).
We can calculate the last U(1) coset branching function to be the extended U(1) affine character
\begin{align}
    \chi^{(l_1,\, l_2;\, l_3)}(q)
    &=\chi^{(Nk(N+k)^2(N+k+1))}_{l_3-(N+k+1)l_1}(q) \nonumber\\   \label{eq:u1branching}
    &=\frac{1}{\eta(q)}\sum_{m\in\, \mathbb{Z}+(l_3-(N+k+1)l_1)/Nk(N+k)^2(N+k+1)}e^{\tfrac{1}{2}Nk(N+k)^2(N+k+1)m^2}\, ,
\end{align}
with the constraint
\begin{align}\label{}
    l_3=l_1+(N+k)l_2 \mod Nk(N+k) ~.
\end{align}
Here $l_3$ is the charge of the $\text{U}(1)_{Nk(N+k)(N+k+1)}$ in the denominator, $l_1$ is the charge of $\text{U}(1)_{Nk(N+k)}$ and $l_2$ the charge of $\text{U}(1)_{Nk}$.
Since we thus know branching functions for the three cosets, we only need to know the branching functions of the three conformal embeddings to get the characters of our new coset.

\subsection{States of the coset theory}

The states of the coset \eqref{eq:KScoset} are labeled by the highest weight representations
as $(\rho, s ; \nu ,m)$. The labels $\rho$ and $\nu$ are highest weights of
su$(N+1)$ and su$(N)$, respectively. The labels $s$ and $m$ take values in $\mathbb{Z}_4$ and
$\mathbb{Z}_{N(N+1)(k+N+1)}$. We are only interested in the NS-sector, so we only use $s=0,2$.
The selection rule is
\begin{align}
 \frac{|\rho |}{N+1} + \frac{s}{2} - \frac{|\nu |}{N} - \frac{m}{N (N+1)} = 0 \text{ mod } 1 ~,
 \label{selection}
\end{align}
where $|\sigma|$ is the number of boxes in the Young tableau corresponding to the weight $\sigma$.
We denote outer automorphisms of $\text{su}(M)$ as $A_{M}$, which are generated
by the cyclic rotations of affine Dynkin labels as in (2.6) of \cite{GG}. Then the field identifications
are
\begin{align}
 (\rho , s ; \nu , m) \simeq ( A_{N+1} \rho , s + 2 ; A_N \nu , m + k + N + 1) ~.
\end{align}
The conformal dimension for the state with $(\rho, s ; \nu ,m)$ is given by
\begin{align}
  h  (\rho , s ; \nu , m)
  = n  + \frac{s}{4} + \frac{1}{ (k + N + 1)} \left( C_{N+1} (\rho ) - C_{N} (\nu) - \frac{m^2}{2N(N+1)} \right)
  \label{hcoset}
\end{align}
for $s=0,2$. Here $C_M (\sigma)$ is the second Casimir operator of su$(M)$ in the representation $\sigma$.
It may be useful to use $C_M (\text{f}) = \frac{M^2 -1}{2M}$ and
$C_M (\text{adj}) = M$ for fundamental and adjoint representations, respectively.
The integer $n$ is the grade at which $(\nu \oplus m)$ appears in
$(\rho \oplus s)$.
In the 't Hooft limit  $N,k \to \infty$ with keeping
\begin{align}
 \lambda = \frac{N}{k+N}
\end{align}
finite, the Casimir eigenvalues become
\begin{equation}
C_N(\mu) \ \sim \ \frac{N|\mu|}{2}\, .
\end{equation}
Thus the conformal dimensions are
\begin{align}
  h  (\rho , s ; \nu , m)
 \ \sim \  n  + \frac{s}{4} +\lambda \frac{|\rho|-|\nu|}{2}- \frac{1}{ (k + N + 1)}\frac{m^2}{2N(N+1)}\, .
  \label{hcosetlimit}
\end{align}

We list conformal dimensions of some simple states.
First let us take $\rho= [1,0^{N}] = \text{f}$ and $\nu = [0^N] = 0$.
Then the selection rule \eqref{selection} may lead to $m = N$.
The conformal weight in the limit is
\begin{align}
  h  (\text{f} , s ; 0 , N) \sim \frac{s}{4} + \frac{\lambda}{2} ~,
\end{align}
where we set $n=0$.
We can have the same states with
conformal weight $\bar h$ for the anti-chiral sector.
The full states consist of the product of chiral and anti-chiral
sectors, and their conformal dimension is $\Delta = h + \bar h$.
These are consistent with the second choice in \eqref{dualcd}.
Next we take  $\rho= 0$ and $\nu = \text{f}$, then
we may choose $m = - N - 1 $. For $s=0$, we
set $n=1$ and then find
\begin{align}
h  ( 0 , 0 ; \text{f} , - N - 1 ) \sim 1 - \frac{\lambda}{2} ~.
\end{align}
For $s=2$, we set $n=0$ and find
\begin{align}
h  ( 0 , 2 ; \text{f} , - N - 1 ) \sim \frac{1}{2} - \frac{\lambda}{2} ~.
\end{align}
These are consistent with the first choice in \eqref{dualcd}.

In analogy to \cite{GG}, this suggests that the characters of the fusion orbits of $(\text{f} , 0 ; 0 , N)$ and $(\text{f} , 2 ; 0 , N)$
together with their conjugate states generate $Z_\text{scalar}^{\frac{\lambda}{2}}Z_\text{scalar}^{\frac{1+\lambda}{2}}\bigl(Z_\text{spinor}^{\frac{1+\lambda}{2}}\bigr)^2$,
but with appropriate glueing of chiral and anti-chiral sectors.
The analogous statement for the characters of  the fusion orbits of $( 0 , 0 ; \text{f} , - N - 1 )$ and $( 0 , 2 ; \text{f} , - N - 1 )$
is that together with their conjugate states they generate $Z_\text{scalar}^{\frac{1-\lambda}{2}}Z_\text{scalar}^{\frac{2-\lambda}{2}}\bigl(Z_\text{spinor}^{\frac{2-\lambda}{2}}\bigr)^2$,
again with appropriate glueing of chiral and anti-chiral sectors.

Let us also consider the second coset description that we introduced by a level-rank duality in eq. \eqref{eq:2ndcoset}. In the same way as above we label the primaries by $(\rho_0,s;\rho_1,\rho_2,q)$.
The conformal weights take the form \cite{Naculich:1997ic}
\begin{align}\label{}
  h  (\rho_0,s;\rho_1,\rho_2,q)
  =&  \frac{s}{4} + \frac{1}{ (k + N + 1)} \left( C_{N+k} (\rho_0 ) - C_{k} (\rho_1)-C_{N} (\rho_2) - \frac{q^2}{2Nk(N+k)} \right)\nonumber \\
   \ \sim& \  \frac{s}{4}+\frac{|\rho_0|}{2} -\frac{1-\lambda}{2} |\rho_1|-\frac{\lambda}{2}|\rho_2|- \frac{q^2}{2Nk(N+k)(N + k + 1)}\, .
\end{align}
Further we have two selection rules in the NS-sector \cite{Lerche:1989uy}
\begin{align}\label{}
    q&=-k|\rho_0|+(k+N)|\rho_1| \mod k(k+N) ~, \nonumber \\
    q&=N|\rho_0|-(k+N)|\rho_2| \mod N(N+k) ~.
\end{align}
A suggestion for the generating states are thus the four states with dimensions
\begin{align}\label{}
    h(\text{f},0;\text{f},0,N)=& \frac{\lambda}{2}\ , & h(\text{f},0;0,\text{f},-k)=&\frac{1-\lambda}{2}\ , \nonumber \\
    h(\text{f},2;\text{f},0,N)=& \frac{1+\lambda}{2}\ , & h(\text{f},2;0,\text{f},-k)=&\frac{2-\lambda}{2}\ .
\end{align}
Note that in this case there is no need for considering heights of the embeddings.
One can also determine the superconformal U(1) charges as \cite{Naculich:1997ic}
\begin{align}\label{}
    Q(\rho_0,s;\rho_1,\rho_2,q)=\frac{s}{2}-\frac{q}{N+k+1} \mod 2 \, ,
\end{align}
and we thus get modulo two
\begin{align}\label{}
        Q(\text{f},0;\text{f},0,N)=& -\lambda\ , & Q(\text{f},0;0,\text{f},-k)=&-(1-\lambda)\ , \nonumber \\
    Q(\text{f},2;\text{f},0,N)=& -(1+\lambda)\ , & Q(\text{f},2;0,\text{f},-k)=&-(2-\lambda)\ ,
\end{align}
that is $-2h$.
A concrete proposal for how to glue the chiral and anti-chiral states and generate the bulk spectrum is that the bosonic matter is generated from
\begin{align}\label{eq:glueingbosonic}
    (\text{f},0;\text{f},0,N)&\otimes(\text{f},0;\text{f},0,N)\ , & (\text{f},0;0,\text{f},-k)&\otimes (\text{f},0;0,\text{f},-k)\ , \nonumber \\
    (\text{f},2;\text{f},0,N)&\otimes (\text{f},2;\text{f},0,N)\ , & (\text{f},2;0,\text{f},-k)&\otimes (\text{f},2;0,\text{f},-k)\ .
\end{align}
and their fusions, and the fermionic matter from the fusions of
\begin{align}\label{}
        (\text{f},0;\text{f},0,N)&\otimes(\text{f},2;\text{f},0,N)\ , & (\text{f},0;0,\text{f},-k)&\otimes(\text{f},2;0,\text{f},-k)\ , \nonumber \\
    (\text{f},2;\text{f},0,N)&\otimes(\text{f},0;\text{f},0,N)\ , & (\text{f},2;0,\text{f},-k)&\otimes(\text{f},0;0,\text{f},-k)\ .
\end{align}

Using the glueing \eqref{eq:glue2} we can calculate the characters corresponding to each of the states. Since the conformal embedding of the Dirac fermions into Majorana fermions contains all orders of even/odd antisymmetric products of the fundamental and antifundamental this will in principle be an infinite series. Calculating the first term containing only the fundamental or the trivial representation gives us:
\begin{align}\label{}
    \chi^{(\text{f},\, 0;\, \text{f},\, 0,\, N)}=& \chi_{1-\lambda}^{(0,\, \text{f};\, \text{f})}\chi_{\lambda}^{(0,\, 0;\, 0)}\chi^{(N,\, 0;\, N)}+\ldots\ , & \chi^{(\text{f},\, 0;\, 0,\text{f},-k)}=&\chi_{1-\lambda}^{(0,\, 0;\, 0)}\chi_{\lambda}^{(0,\, \text{f};\, \text{f})}\chi^{(-k,\, 0;\, -k)}+\ldots\ , \nonumber \\
    \chi^{(\text{f},\, 2;\, \text{f},\, 0,\, N)}=&\chi_{1-\lambda}^{(\mathrm{f},\, 0;\text{f})}\chi_{\lambda}^{(\bar{\mathrm{f}},\, \text{f};\, 0)}\chi^{(-k,\, 1;\, N)}+\ldots \ , & \chi^{(\text{f},\, 2;\, 0,\, \text{f},\, -k)}=&\chi_{1-\lambda}^{(\bar{\mathrm{f}},\, \text{f};\, 0)}\chi_{\lambda}^{(\text{f},\, 0;\, \text{f})}\chi^{(N,\, -1;\, -k)}+\ldots\ .
\end{align}
Here $\chi_{\lambda}^{(\rho_1,\, \rho_2;\, \rho_3)}$ is the branching function of the large $N$ minimal coset, and $\chi^{(l_1,\, l_2;\, l_3)}$ is the branching function for the U(1) coset \eqref{eq:u1branching}. Each of our basic characters reduces to a single minimal coset character times the vacuum part for the remaining two characters which have vanishing conformal dimension. Note that the bosonic part \eqref{eq:glueingbosonic} indeed correspond to the generators of the expected minimal cosets \eqref{eqrelationtoGG} (after considering also the conjugated states) with the spin one part coming from the U(1) branching function.
The branching functions have been computed in \cite{GG,GGHR}. This allows us to compute some leading terms of the CFT partition function.
We get
\begin{equation}
\begin{split}
Z_{\text{CFT}}\ &= \ 1+q+\bar q + 2\bigl(1+2q+2\bar q+...)\Bigl((q\bar q)^{\frac{\lambda}{2}}+(q\bar q)^{\frac{1-\lambda}{2}}+(q\bar q)^{\frac{1+\lambda}{2}}+(q\bar q)^{\frac{2-\lambda}{2}}+\\
&\qquad +q^{\frac{\lambda}{2}}\bar q^{\frac{1+\lambda}{2}}+q^{\frac{1-\lambda}{2}}\bar q^{\frac{2-\lambda}{2}}+q^{\frac{1+\lambda}{2}}\bar q^{\frac{\lambda}{2}}+q^{\frac{2-\lambda}{2}}\bar q^{\frac{1-\lambda}{2}}+...\Bigl)
\end{split}
\end{equation}
This indeed agrees with leading terms of the bulk partition function.

\section{Conclusions and outlook}

In this paper, we proposed that Vasiliev's higher spin supergravity theory with symmetry algebra shs$[\lambda]$ \cite{PV1,PV2} is the holographic dual
to the $\mathcal N=(2,2)$ $\mathbb C$P$^N$ Kazama-Suzuki coset \cite{KS}
\begin{align}
 \frac{\text{SU}(N+1)_k \times \text{SO}(2 N)_1 }{\text{SU}(N)_{k+1} \times \text{U}(1)_{N(N+1)(k+N+1)} }
\end{align}
in the 't Hooft limit with $N,k$ infinite, but
\begin{equation}
\lambda \ = \ \frac{N}{k+N}
\end{equation}
fixed.

The conjecture is based on symmetry and partition function considerations.
The symmetry of the CFT is the $\mathcal N=2$ ${\cal W}_{N+1}$ algebra.
This algebra is obtained via Drinfeld-Sokolov reduction from affine sl$(N+1|N)$
corresponding to a certain principal embedding of sl(2).
The massless sector of higher spin AdS$_3$ supergravity is a large $N$ limit of
SL$(N+1|N)$ $\otimes$ SL$(N+1|N)$ Chern-Simons theory.
Imposing AdS boundary conditions is equivalent to the Drinfeld-Sokolov reduction of the corresponding
current algebra \cite{CFPT,CFP}. Thus the asymptotic symmetry near the boundary
is a large $N$ limit of  $\mathcal N=2$ ${\cal W}_{N+1}$ algebra.
We compute the bulk partition function and compare it with the CFT.
In particular, the computation of one loop determinant for the higher spin
spinors is new.
The vacuum character of the CFT agrees with the massless part of the higher spin theory.
The level-rank duality and some supersymmetry considerations further support our conjecture
that the bulk partition function agrees with the one of the CFT. For the matter sector we could compute the characters to lowest non-trivial order and find agreement with the bulk.

We also observe that the partition function is invariant under the exchange of couplings
\begin{equation}
\lambda\ \longleftrightarrow\ 1-\lambda\ .
\end{equation}
On the CFT side this exchange of coupling constants is a level-rank duality.
It thus seems likely that the theories possess a strong-weak self-duality.
It should be verified in what sense the self-duality holds.

In order to establish the duality proposed in this paper, we have to
collect more evidence. The supergravity partition function is computed as in
\eqref{eq:totalbulk}, and this should be reproduced by taking the 't Hooft
limit of the CFT partition function.
We have already obtained several supports, but the
direct proof is desired.
Even if it is difficult, the comparison of elliptic genus would give some
important clues.
The conjecture is also based on the
symmetry argument. For the Chern-Simons theory with $\text{sl}(N+1|N)$ symmetry,
we have seen that the asymptotic symmetry is ${\cal N}=2$ ${\cal W}_{N+1}$ algebra.
However, Vasiliev's theory has infinite dimensional symmetry shs$[\lambda]$,
and we need to take a large $N$ limit with a special care
for the analysis of asymptotic symmetry.
Another possible test of the duality is to compare the RG-flow of both theories.
Since the theories of the both sides of duality are tractable ones,
there may be a chance to give a proof of the duality including the bosonic
sub-sector.

Once the large $N$ equivalence is well-established, we may include
$1/N$ corrections.
For the CFT side it is a trivial task, but for the gravity side
it means that we have to deal with the quantum corrections of the gravity theory.
Speaking in a different way, this duality would give a hint to quantize the gravity theory.
We hope that the extension to supersymmetric models helps us for
this investigation.

\subsection*{Acknowledgements}

We are grateful to Thomas Quella, Soo-Jong Rey and Tadashi Takayanagi for useful
discussions. YH would like to thank IMPU for its hospitality, where a part
of this work was done. The work of YH is supported in part by Keio Gijuku Academic Development Funds, and the work of PBR is funded by DFG grant no. ZI 513/2-1.

\appendix

\section{Higher spin algebras }
\label{hsa}

In this appendix, we correct some useful facts on the infinite dimensional algebras
hs$[\lambda]$ and shs$[\lambda]$.

\subsection{Higher spin algebra hs[$\lambda$] }

Let us first review the bosonic hs[$\lambda$] algebra.
This algebra includes generators $V_n^s$ with $s \geq 2$ and $|n| < s$.
Among them $V^2_0,V^2_{\pm 1}$ generate sl(2) subalgebra and satisfies
\begin{align}
 [ V^2_m , V^s_n ] = ( - n + m (s - 1) ) V^s_{m+n} ~.
\end{align}
The other commutation relations are
\begin{align}
 [ V_m^s , V_n^t] = \sum_{l}^{[(s+t-1)/2]} g_{2l}^{st} (m,n;\lambda) V_{m+n}^{s+t-2l} \label{hssc}
\end{align}
where $[a]$ is the maximal integer number less than $a$. The structure constants
are given as \cite{PRS,GH,CFP}
\begin{align}
 g_{u}^{st} (m,n;\lambda) = \frac{1}{2 (u-1)!} \phi_u^{st} (0 , \lambda ) N_u^{st} (m,n)
\end{align}
with
\begin{align} \nonumber
& N_u^{st} (m,n) = \sum_{k=0}^{u-1} (-1)^k \begin{pmatrix} u-1 \\ k \end{pmatrix}
 [ s-1+m]_{u-1+k } [s - 1 - m]_k [t - 1 + n]_k [t - 1 - n]_{u-1-k} ~, \\
& \phi_u^{st} (\rho , \lambda)
 = {}_4 F_3 \left[ \begin{matrix} \frac{1}{2} + \lambda - \rho , \frac{1}{2} - \lambda - \rho ,
     \frac{2 - u}{2} + \rho ,   \frac{1 - u}{2} + \rho \\
  \frac{3}{2} - s , \frac{3}{2} - t , \frac{1}{2} + s + t - u \end{matrix} \right| 1 \biggr]~.
  \label{Nphi}
\end{align}
Here $[a]_n = \Gamma(a+1)/\Gamma(a+1-n)$.
There are several ways to describe the algebra. Some of them are
\begin{itemize}
\item[(i)] the quotient of the universal enveloping algebra $U$(sl(2)) by the ideal generated
 by $(C_2 - \mu \bf{1})$, say \cite{PRS,GH}. Here $C_2$ is the Casimir operator of sl(2).
\item[(ii)] analytic continuation of SU($N$) into $N=\nu$ with $\nu \in \mathbb{R}$ \cite{FL}.
\item[(iii)] algebra generated by differential operators on the circle $S^1$, say \cite{FL}.
\end{itemize}

\subsection{Higher spin superalgebra shs[$\lambda$]}

We would like to construct ${\cal N}=2$ superalgebra shs[$\lambda$] as in the previous subsection.
The three methods are now
\begin{itemize}
\item[(i)] the quotient of the universal enveloping algebra $U$(osp$(1|2)$) by the ideal generated
 by $(C_2 - \mu \bf{1})$ \cite{BWV,Vasiliev}. Here $C_2$ is the Casimir operator of osp$(1|2)$.
\item[(ii)] analytic continuation of SU($N+1|N$) into $N=\nu$ with $\nu \in \mathbb{R}$ \cite{FL}.
\item[(iii)] algebra generated by differential operators with super-coordinate $(z,\theta)$ \cite{BVW,BWV}.
\end{itemize}
The definition of superalgebra used in the Vasiliev's theory \cite{PV1,PV2}
is (i) in the language of star product, and it was shown in \cite{BWV} that the
superalgebra is the same as the one in (iii).

The superalgebra in \cite{BVW,BWV} includes osp$(1|2)$ subalgebra, whose generators
$V^{(2)+}_0$, $V^{(2)+}_{\pm 1}$, $F^{(1)+}_{\pm1/2}$ satisfy
\begin{align}
& [V^{(2)+}_m , V^{(2)+}_n] = (m-n) V_{m+n}^{(2)+}~,  \qquad
 [V^{(2)+}_m , F^{(1)+}_r] = (\tfrac{1}{2} m - r) F^{(1)+}_{m+r} ~, \nonumber \\
& \{ F^{(1)+}_r, F^{(1)+}_s\} = 2 V^{(2)+}_{r+s} ~.
\end{align}
Among the other generators, (anti-)commutation relations are
\begin{align}
&[V_m^{(2)+} , V_n^{(s)\pm}] = (- n + m (s-1) ) V_{m+n}^{(s)\pm} ~, \qquad
[V_m^{(2)+} , F_r^{(s)\pm}] = (- r + m (s-\tfrac{1}{2}) )F_{m+n}^{(s)\pm} ~, \nonumber \\
&[F_{1/2}^{(1)+} , V_m^{(s)+}] = - \tfrac12 (m -s +1) F_{m+1/2}^{(s-1)+} ~, \qquad
[F_{1/2}^{(1)+} , V_m^{(s)+}] = - F_{m+1/2}^{(s)+} ~, \\
&\{ F^{(1)+}_{1/2} , F_r^{(s-1)+}\} = 2 V_{r+1/2}^{(s)+} ~, \qquad
\{ F^{(1)+}_{1/2} , F^{(s)-}_r \} = (r-s +\tfrac{1}{2}) V_{r+1/2}^{(s)-}  ~.  \nonumber
\end{align}
It was argued that $(V^{(1)-}_0 , F^{(1)\pm}_{\pm 1/2} , V^{(2)+}_0 , V^{(2)+}_{\pm 1} )$
generate osp$(2|2)$ subalgebra.
Here the labels take $n \in \mathbb{Z}$ and $r \in \mathbb{Z} + 1/2$
satisfying $|n| \leq s - 1$ and $|r| \leq s - 1/2  $.
The other commutation relations can be found in \cite{BWV} in principle.
For $\lambda = 1$ the structure constants are given in \cite{BPRSS} in terms of
functions \eqref{Nphi}, therefore it is natural to expect that they are written in
a similar way even for generic $\lambda$. Probably they are also found in \cite{FL}
as for SU$(\nu+1|\nu)$, see (ii) above.

\section{The heat kernel method}
\label{heat}

Applying the heat kernel method as in \cite{GMY,DGG},
we compute  \eqref{hspinor} and \eqref{hsb} with $s=1$.
We use the formula in (6.9) and (7.2) of \cite{DGG}
\begin{align}
 - \log \text{det} (- \Delta_{(s)} + m_s^2)
 =
 \int_0^\infty \frac{dt}{t} K^{(s)} (\tau , \bar \tau ; t) e^{-m_s^2 t}
 \label{hkdet}
\end{align}
with
\begin{align}
 K^{(s)} (\tau , \bar \tau ; t)
  =  (2 - \delta_{s,0} )
\sum_{m=1}^{\infty} \frac{ (-1)^{2 s m} \tau_2}{4 \sqrt{\pi t}
  |\sin \frac{m \tau}{2}|^2} \cos (s m \tau_1) e^{- \frac{m^2 \tau_2^2}{4 t}}
  e^{- (s+1)t}
\end{align}
for relevant parts.
Here the phase factor $(-1)^{2 s m}$ is included to explain the
anti-periodicity of half-integer spin particle.

\subsection{Massive spin $1/2$ fermions}

For a Dirac fermion with spin $1/2$ and mass $M$,
the partition function at the one loop level is
\begin{align}
Z^h_\text{spinor} =  \text{det} (\nabla + M)_{(\frac12)}
 =  \text{det} ^{\frac12} (- \Delta - \tfrac{3}{2} + M^2  )_{(\frac12)} ~.
\end{align}
The shift of mass is due to the AdS curvature as in \eqref{shiftmass}.
Here the relation to the dual conformal dimension is
$\Delta_F = 2h -1 = 1 + M$. Using the integral formula
\begin{align}
 \frac{1}{4 \pi^{1/2}} \int_0^\infty \frac{dt}{t^{3/2}}
 e^{- \frac{\alpha^2}{4t} - \beta ^2 t} = \frac{1}{2 \alpha} e^{- \alpha \beta} ~,
\end{align}
we find
\begin{align}
 \log Z^h_\text{spinor} = - \frac12 \sum_{m=1}^\infty \frac{(-1)^m}{m |\sin \frac{m \tau}{2}|}
   \cos \left( \tfrac{m}{2} \tau_1 \right)  e^{- m \tau_2 M }
   =  -  \sum_{m=1}^\infty \frac{(-1)^m|q^m|^{2h - 1}}{m|1 - q^m|^2}
 (q^{\frac{m}{2}} + \bar q^{\frac{m}{2}}) ~.
\end{align}
Therefore, the one loop determinant is
\begin{align}
 Z^h_\text{spinor} &=
 \exp \left( - \sum_{m=1}^\infty \frac{(-1)^m|q^m|^{2h - 1}}{m|1 - q^m|^2}
 (q^{\frac{m}{2}} + \bar q^{\frac{m}{2}}) \right) \nonumber  \\
  &= \exp \left( - \sum_{m=1}^\infty \sum_{l,l'=0}^\infty
 \frac{(-1)^m}{m}
 (q^{m (h + l)} \bar q^{m (h - \frac12 + l')}
 + q^{m (h - \frac12 + l)} \bar q^{m (h + l')}) \right) \\
 &= \prod_{l,l' = 0}^\infty ( 1 + q^{h + l } \bar q^{h - \frac12 + l'} )
  ( 1 + q^{h - \frac12 + l } \bar q^{h  + l'} )  \nonumber
\end{align}
as in \eqref{hspinor}.

\subsection{Vector fields}

The $s=1$ sub-sector in the decomposition \eqref{slexp} of $\text{SL}(N+1|N)$
supergroup is given by the Abelian group, so the Chern-Simons action
at the free limit becomes quite simple as
\begin{align}
 S \propto  \int d ^3 x \epsilon^{\mu \nu \rho}
 ( a_\mu \nabla_\nu a_\rho - \tilde a_\mu \nabla_\nu \tilde a_\rho )
  = \int d ^3 x \epsilon^{\mu \nu \rho}
  ( a_\mu + \tilde a_\mu ) \nabla_\nu ( a_\rho - \tilde a_\rho ) ~.
  \label{aaction}
\end{align}
We have used $\nabla_\mu$ for the covariant derivative on the
AdS space. The gauge symmetry is generated by
\begin{align}
 \delta a_\mu = \nabla_\mu \lambda ~, \qquad
 \delta \tilde a_\mu = \nabla_\mu \tilde \lambda ~.
\end{align}
Notice that there is no contribution from $A \wedge A \wedge A$ like terms for
the Abelian sector. If such terms exist, then we can integrate
$\omega_\mu = \frac12 (A_\mu + \tilde A_\mu )$ and obtain the free action
for $e_\mu = \frac12 (A_\mu - \tilde A_\mu)$ used
in \cite{GGS}. The action used in \cite{GGS} with $s=1$ is Yang-Mills action in three
dimension, but we cannot use it for the reason. In fact, our theory is topological,
but Yang-Mills theory is not. For Yang-Mills theory, the one loop determinant
was computed in \cite{GMY}.

As in \cite{GGS} we decompose the gauge field as
\begin{align}
 a_\mu = a^\text{T}_\mu + \nabla_\mu \lambda ~, \qquad
 \tilde a_\mu = \tilde a^\text{T}_\mu + \nabla_\mu \tilde \lambda ~, \qquad
 \nabla^\mu  a^\text{T}_\mu = \nabla^\mu \tilde a^\text{T}_\mu = 0 ~.
\end{align}
If we put this decomposition into the action \eqref{action},
then we obtain the same expression with $a_\mu , \tilde a_\mu$
replaced by $a^\text{T}_\mu , \tilde a^\text{T}_\mu$.
We fix the gauge by setting $\lambda = \tilde \lambda = 0$,
which leads to the Faddeev-Popov determinant as
\begin{align}
  ( \text{det} ^{\frac12} ( - \Delta )_{(0)} )^2 ~.
\end{align}
The subscript implies that the derivative acts on a scalar field.

For the transverse components, the one loop determinant is given by
\begin{align}
 \text{det}^{-1} (\epsilon^{\mu \rho \nu} \nabla_\rho)^\text{T}_{(1)} ~.
\end{align}
Denoting $e_\mu^\text{T} = \frac12 (a_\mu^\text{T} + \tilde a_\mu^\text{T})$, we have
\begin{align}
 (\epsilon^{\mu \rho \nu} \nabla_\rho \epsilon^{\nu \sigma \delta} \nabla_\sigma) e^\text{T}_\delta
 &= (g^{\mu \sigma} g^{\rho \delta} - g^{\mu \delta} g^{\rho \sigma})
 \nabla_\rho \nabla_\sigma e^\text{T}_\delta \\
 &= ( [\nabla^\delta , \nabla^\mu] - g^{\mu \delta} \Delta ) e^\text{T}_\delta
 =( - \Delta - 2) e^{\text{T}\mu} ~. \nonumber
\end{align}
In the last equality we have used the formula \eqref{formula}.
Therefore, the determinant can be written as
\begin{align}
 \text{det}^{-1} (\epsilon^{\mu \nu \rho} \nabla_\mu)^\text{T}_{(1)}
   = \text{det}^{-\frac12} ( - \Delta - 2)^\text{T}_{(1)} ~.
\end{align}

In total, the one loop determinant is given by
\begin{align}
Z_B^{(1)} = \frac{ \text{det} ( - \Delta )_{(0)} }{\text{det}^{\frac12} ( - \Delta - 2)^\text{T}_{(1)}}
 ~.
\end{align}
Applying the formula \eqref{hkdet}, we have
\begin{align}
\log Z^{(1)}_B &=- \frac12 \sum_{m=1}^\infty \frac{1}{m|\sin \frac{m \tau}{2}|^2}
 \left[ \cos (m \tau_1)
 -     e^{ - m \tau_2 } \right]
 \nonumber \\
  &= - \sum_{m=1}^\infty \frac{1}{m} \left[ \frac{q^{m }}{1 - q^m}
 + \frac{\bar q^{m }}{1 - \bar q^m} \right]
 = - \log \left[ \prod_{n=1}^\infty | 1 - q^{n} |^2 \right] ~.
\end{align}
Therefore, we obtain
\begin{align}
Z^{(1)}_B
 = \prod_{n=1}^\infty \frac{1}{ | 1 - q^{n} |^2 }
\end{align}
as given in \eqref{hsb} with $s=1$.


\begin{thebibliography}{9}

\bibitem{GG}
  M.~R.~Gaberdiel, R.~Gopakumar,
  ``An AdS$_3$ dual for minimal model CFTs,''
  Phys.\ Rev.\  {\bf D83}, 066007 (2011).
  [arXiv:1011.2986 [hep-th]].


\bibitem{PV1}
  S.~F.~Prokushkin, M.~A.~Vasiliev,
  ``Higher spin gauge interactions for massive matter fields in 3D AdS space-time,''
  Nucl.\ Phys.\  {\bf B545}, 385 (1999).
  [arXiv:hep-th/9806236 [hep-th]].

\bibitem{PV2}
  S.~Prokushkin, M.~A.~Vasiliev,
  ``3d higher spin gauge theories with matter,''
  [hep-th/9812242].


\bibitem{KP}
  I.~R.~Klebanov, A.~M.~Polyakov,
  ``AdS dual of the critical O$(N)$ vector model,''
  Phys.\ Lett.\  {\bf B550}, 213-219 (2002).
  [hep-th/0210114].

\bibitem{Vasiliev:2003ev}
  M.~A.~Vasiliev,
  ``Nonlinear equations for symmetric massless higher spin fields in (A)dS$_d$,''
  Phys.\ Lett.\  {\bf B567}, 139-151 (2003).
  [hep-th/0304049].


\bibitem{Maldacena:1997re}
  J.~M.~Maldacena,
  ``The Large $N$ limit of superconformal field theories and supergravity,''
  Adv.\ Theor.\ Math.\ Phys.\  {\bf 2}, 231-252 (1998).
  [hep-th/9711200].

\bibitem{Blencowe}
  M.~P.~Blencowe,
  ``A consistent interacting massless higher spin field theory in $D = (2+1)$,''
  Class.\ Quant.\ Grav.\  {\bf 6}, 443 (1989).

\bibitem{HR}
  M.~Henneaux, S.~-J.~Rey,
  ``Nonlinear ${\cal W}_{\infty}$ as asymptotic symmetry of three-dimensional higher spin anti-de Sitter gravity,''
  JHEP {\bf 1012}, 007 (2010).
  [arXiv:1008.4579 [hep-th]].


\bibitem{CFPT}
  A.~Campoleoni, S.~Fredenhagen, S.~Pfenninger, S.~Theisen,
  ``Asymptotic symmetries of three-dimensional gravity coupled to higher-spin fields,''
  JHEP {\bf 1011}, 007 (2010).
  [arXiv:1008.4744 [hep-th]].

\bibitem{GH}
  M.~R.~Gaberdiel, T.~Hartman,
  ``Symmetries of holographic minimal models,''
  JHEP {\bf 1105}, 031 (2011).
  [arXiv:1101.2910 [hep-th]].


\bibitem{CFP}
  A.~Campoleoni, S.~Fredenhagen, S.~Pfenninger,
  ``Asymptotic ${\cal W}$-symmetries in three-dimensional higher-spin gauge theories,''
  [arXiv:1107.0290 [hep-th]].

\bibitem{GGS}
  M.~R.~Gaberdiel, R.~Gopakumar, A.~Saha,
  ``Quantum ${\cal W}$-symmetry in AdS$_3$,''
  JHEP {\bf 1102}, 004 (2011).
  [arXiv:1009.6087 [hep-th]].

\bibitem{GGHR}
  M.~R.~Gaberdiel, R.~Gopakumar, T.~Hartman, S.~Raju,
  ``Partition functions of holographic minimal models,''
  JHEP {\bf 1108}, 077 (2011).
  [arXiv:1106.1897 [hep-th]].

\bibitem{Chang:2011mz}
  C.~-M.~Chang, X.~Yin,
  ``Higher spin gravity with matter in AdS$_3$ and its CFT Dual,''
  [arXiv:1106.2580 [hep-th]].

\bibitem{Papadodimas:2011pf}
  K.~Papadodimas, S.~Raju,
  ``Correlation functions in holographic minimal models,''
  [arXiv:1108.3077 [hep-th]].

\bibitem{Ahn:2011by}
  C.~Ahn,
  ``The coset spin-4 casimir operator and its three-point functions with scalars,''
  [arXiv:1111.0091 [hep-th]].

\bibitem{Kiritsis:2010xc}
  E.~Kiritsis, V.~Niarchos,
  ``Large-$N$ limits of 2d CFTs, quivers and AdS$_3$ duals,''
  JHEP {\bf 1104}, 113 (2011).
  [arXiv:1011.5900 [hep-th]].

\bibitem{Ahn}
  C.~Ahn,
  ``The large $N$ 't Hooft limit of coset minimal models,''
  [arXiv:1106.0351 [hep-th]].

\bibitem{Gaberdiel:2011nt}

  M.~R.~Gaberdiel, C.~Vollenweider,
  ``Minimal model holography for SO$(2N)$,''
  [arXiv:1106.2634 [hep-th]].

\bibitem{Ouyang:2011fs}
  P.~Ouyang,
  ``Toward higher spin dS$_3/$CFT$_2$,''
  [arXiv:1111.0276 [hep-th]].

\bibitem{Ising}
  A.~Gastro, M.~R.~Gaberdiel, T.~Hartman, A.~Maloney, R.~Volpato,
  ``The gravity dual of the Ising model,''
  [arXiv:1111.1987 [hep-th]].

\bibitem{AT}
  A.~Achucarro, P.~K.~Townsend,
  ``A Chern-Simons action for three-dimensional anti-de Sitter supergravity theories,''
  Phys.\ Lett.\  {\bf B180}, 89 (1986).

\bibitem{HMS}
  M.~Henneaux, L.~Maoz, A.~Schwimmer,
  ``Asymptotic dynamics and asymptotic symmetries of three-dimensional extended AdS supergravity,''
  Annals Phys.\  {\bf 282}, 31-66 (2000).
  [hep-th/9910013].


\bibitem{KS}
  Y.~Kazama, H.~Suzuki,
  ``New ${\cal N}=2$ superconformal field theories and superstring compactification,''
  Nucl.\ Phys.\  {\bf B321}, 232 (1989).

\bibitem{Naculich:1997ic}
  S.~G.~Naculich, H.~J.~Schnitzer,
  ``Superconformal coset equivalence from level rank duality,''
  Nucl.\ Phys.\  {\bf B505}, 727-748 (1997).
  [hep-th/9705149].

\bibitem{GMY}
  S.~Giombi, A.~Maloney, X.~Yin,
  ``One-loop partition functions of 3D gravity,''
  JHEP {\bf 0808}, 007 (2008).
  [arXiv:0804.1773 [hep-th]].

\bibitem{DGG}
  J.~R.~David, M.~RGaberdiel, R.~Gopakumar,
  ``The heat kernel on AdS$_3$ and its applications,''
  JHEP {\bf 1004}, 125 (2010).
  [arXiv:0911.5085 [hep-th]].

\bibitem{FL}
  E.~S.~Fradkin, V.~Y.~.Linetsky,
  ``Supersymmetric Racah basis, family of infinite dimensional superalgebras, SU$(\infty + 1|\infty)$ and related 2D models,''
  Mod.\ Phys.\ Lett.\  {\bf A6}, 617-633 (1991).

\bibitem{W-alg}
  P.~Bouwknegt, K.~Schoutens,
  ``${\cal W}$ symmetry in conformal field theory,''
  Phys.\ Rept.\  {\bf 223}, 183-276 (1993).
  [hep-th/9210010].

\bibitem{FronsdaldS}
  C.~Fronsdal,
  ``Singletons and massless, integral spin fields on de Sitter space (Elementary particles in a curved space. 7.,''
  Phys.\ Rev.\  {\bf D20}, 848-856 (1979).

\bibitem{Campoleoni}
  A.~Campoleoni,
  ``Metric-like Lagrangian formulations for higher-spin fields of mixed symmetry,''
  Riv.\ Nuovo Cim.\  {\bf 033}, 123-253 (2010).
  [arXiv:0910.3155 [hep-th]].

\bibitem{MW}
  A.~Maloney, E.~Witten,
  ``Quantum gravity partition functions in three dimensions,''
  JHEP {\bf 1002}, 029 (2010).
  [arXiv:0712.0155 [hep-th]].

\bibitem{KW}
  I.~R.~Klebanov, E.~Witten,
  ``AdS/CFT correspondence and symmetry breaking,''
  Nucl.\ Phys.\  {\bf B556}, 89-114 (1999).
  [hep-th/9905104].

\bibitem{FFdS}
  J.~Fang, C.~Fronsdal,
  ``Massless, half integer spin fields in de Sitter space,''
  Phys.\ Rev.\  {\bf D22}, 1361 (1980).

\bibitem{dWF}
  B.~de Wit, D.~Z.~Freedman,
  ``Systematics of higher spin gauge fields,''
  Phys.\ Rev.\  {\bf D21}, 358 (1980).

\bibitem{FF}
  J.~Fang, C.~Fronsdal,
  ``Massless fields with half integral spin,''
  Phys.\ Rev.\  {\bf D18}, 3630 (1978).

\bibitem{Fronsdal}
  C.~Fronsdal,
  ``Massless fields with integer spin,''
  Phys.\ Rev.\  {\bf D18}, 3624 (1978).

\bibitem{Nozaki}
  M.~Nozaki,
  ``Comments on D-branes in Kazama-Suzuki models and Landau-Ginzburg theories,''
  JHEP {\bf 0203}, 027 (2002).
  [hep-th/0112221].


\bibitem{Ito}
  K.~Ito,
  ``Quantum Hamiltonian reduction and ${\cal N}=2$ coset models,''
  Phys.\ Lett.\  B {\bf 259}, 73 (1991).

\bibitem{Gepner:1988wi}
  D.~Gepner,
  ``Scalar field theory and string compactification,''
  Nucl.\ Phys.\  {\bf B322 } (1989)  65.


\bibitem{Altschuler:1988mg}
  D.~Altschuler,
  ``Quantum equivalence of coset space models,''
  Nucl.\ Phys.\  {\bf B313}, 293 (1989).

\bibitem{Bowcock:1988vs}
  P.~Bowcock, P.~Goddard,
  ``Coset constructions and extended conformal algebras,''
  Nucl.\ Phys.\  {\bf B305}, 685 (1988).



\bibitem{DiFrancesco:1993dg}
  P.~Di Francesco, S.~Yankielowicz,
  ``Ramond sector characters and ${\cal N}=2$ Landau-Ginzburg models,''
  Nucl.\ Phys.\  {\bf B409}, 186-210 (1993).
  [hep-th/9305037].

\bibitem{Lerche:1989uy}
  W.~Lerche, C.~Vafa, N.~P.~Warner,
  ``Chiral rings in ${\cal N}=2$ superconformal theories,''
  Nucl.\ Phys.\  {\bf B324}, 427 (1989).


\bibitem{PRS}
  C.~N.~Pope, L.~J.~Romans, X.~Shen,
  ``W$_\infty$ and the Racah-Wigner algebra,''
  Nucl.\ Phys.\  {\bf B339}, 191-221 (1990).


\bibitem{BWV}
  E.~Bergshoeff, B.~de Wit, M.~A.~Vasiliev,
  ``The structure of the super-W$_\infty$($\lambda$) algebra,''
  Nucl.\ Phys.\  {\bf B366}, 315-346 (1991).

\bibitem{Vasiliev}
  M.~A.~Vasiliev,
  ``Higher spin gauge theories: Star product and AdS space,''
  In *Shifman, M.A. (ed.): The many faces of the superworld* 533-610.
  [hep-th/9910096].

\bibitem{BVW}
  E.~Bergshoeff, M.~A.~Vasiliev, B.~de Wit,
  ``The super-W$_\infty$($\lambda$) algebra,''
  Phys.\ Lett.\  {\bf B256}, 199-205 (1991).


\bibitem{BPRSS}
  E.~Bergshoeff, C.~N.~Pope, L.~J.~Romans, E.~Sezgin, X.~Shen,
  ``The super W$_\infty$ algebra,''
  Phys.\ Lett.\  {\bf B245}, 447-452 (1990).


\end{thebibliography}
\end{document}